\pdfoutput=1
\documentclass[11pt, a4paper, nonumbering, copyright, trilogo]{google}

\usepackage[numbers,square,comma,sort&compress]{natbib}
\usepackage{graphicx}
\usepackage{booktabs}
\usepackage{tabularx}
\usepackage{longtable}
\usepackage{array}
\usepackage{seqsplit}
\usepackage{pdflscape}
\uselogo{true}
\title{Aptamers as a sandbox for lab-in-the-loop agentic molecular discovery}
\correspondingauthor{pjro@google.com, apawlosky@google.com}
\reportnumber{}

\renewcommand\today{2026-10-12}

\author[1]{Zachary Cutts}
\author[1]{John Coller}
\author[1]{Rachel Chen}
\author[2,$\ddagger$]{Hua Tang}
\author[1]{Mathias Bellaiche}
\author[1]{Ryder Easterlin}
\author[3]{Dina Zverinski}
\author[4]{Petar Sirkovic}
\author[1]{Avinatan Hassidim}
\author[1]{Yossi Matias}
\author[1,$\dagger$]{Pooja Rao}
\author[1,$\dagger$]{Annalisa Pawlosky}

\affil[1]{Google Research}
\affil[2]{Department of Genetics, Stanford University, Stanford, CA, USA}
\affil[3]{Google Cloud}
\affil[4]{Google DeepMind}
\affil[$\dagger$]{Equal leadership (co-senior authors)}
\affil[$\ddagger$]{Work performed while H.T. was a Visiting Faculty Researcher at Google Research}

\begin{abstract}
Despite recent breakthroughs in AI-driven molecular generation, existing discovery pipelines operate largely \textit{in silico}, with wet-lab feedback reserved for final validation rather than iterative improvement. Whether large language model (LLM)-based reasoning agents can meaningfully accelerate \textit{de novo} discovery by iterating with real-world lab data remains unproven.

Here we use protein-binding aptamer design as a high-throughput sandbox (1--2 week validation cycles, \$1--\$3 per candidate) to demonstrate the potential of a wet lab in a closed loop with AI systems. We tasked agentic pipelines with developing aptamer binders against the protein targets $\alpha$-thrombin and myoglobin across three design--test--improve cycles with a budget of $\sim$1,000 candidates per round ($\sim$4,400 total sequenced constructs). Statistically significant, bead-corrected on-target hit rates among newly designed candidates rose across rounds from 0.0\% in Round 1 to 26.0\% in Round 2 and 42.1\% in Round 3 on $\alpha$-thrombin (37.1\% excluding literature motif descendants), and from 10.2\% in Round 1 to 3.7\% in Round 2 and 33.5\% in Round 3 on myoglobin. Orthogonal single-clone bead-binding titrations of eight full-length $\alpha$-thrombin candidates confirmed quantitative concordance with pooled sequencing enrichment (Pearson $r = 0.90$). In these assays, literature-derived motifs engineered by the agents to reduce off-target reactivity in the pooled screen reached preliminary apparent affinities of $K_D = 24.6\text{--}133\text{ nM}$, whereas strictly \textit{de novo} scaffolds showed low-micromolar binding ($K_D > 1\ \mu\text{M}$).

We also report how our pipelines used binding data on myoglobin to initially exploit the high-affinity G-quadruplex binder space, then diagnose cross-reactivity as on-target specificity dropped from 75.0\% to 19.1\%, and subsequently recover by developing new strategies that restored myoglobin specificity to 72.4\%. We demonstrate that empirical lab data is much more valuable than existing computational proxies, showing how iterative experimental feedback enables AI agents to diagnose assay artifacts and refine molecular designs across rounds.
\end{abstract}

\begin{document}
\maketitle

\section*{Introduction}

Multi-agent systems for scientific discovery, often termed "AI co-scientists," have begun to accelerate biological research by generating testable hypotheses, mining sequence databases, and designing candidate molecules [\hyperlink{ref1}{1}, \hyperlink{ref2}{2}, \hyperlink{ref3}{3}, \hyperlink{ref4}{4}]. Although active learning is well established in high-throughput screening [\hyperlink{ref5}{5}, \hyperlink{ref6}{6}], initial studies with large language model (LLM)-based reasoning agents have validated their designs in only a single experimental pass at the end of the workflow [\hyperlink{ref1}{1}, \hyperlink{ref2}{2}, \hyperlink{ref3}{3}, \hyperlink{ref4}{4}]. Extending these systems into multi-round experimental loops, where agents formulate hypotheses, diagnose failures from empirical data, and revise their strategy across successive wet-lab cycles, therefore represents a natural next frontier.

AI agents operate within software harnesses that interleave reasoning with tool execution, code generation, literature retrieval, and persistent memory, allowing them to update their decisions as new data arrive [\hyperlink{ref7}{7}, \hyperlink{ref8}{8}, \hyperlink{ref9}{9}]. As both general-purpose and biology-specific agent harnesses rapidly proliferate [\hyperlink{ref1}{1}, \hyperlink{ref2}{2}, \hyperlink{ref8}{8}, \hyperlink{ref9}{9}], physical wet-lab validation has become the primary bottleneck in biological discovery. In chemistry and materials science, robotic laboratory automation has helped overcome this bottleneck, giving rise to "self-driving laboratories" where AI systems plan, physically execute, and analyze iterative experiments with minimal human intervention [\hyperlink{ref10}{10}, \hyperlink{ref11}{11}]. In biomolecular design, closing the loop is far harder because synthesis, expression, and functional validation remain slow and expensive [\hyperlink{ref12}{12}].

In domains such as games, mathematics, and software engineering, self-play and iterative agentic search succeed because verification is purely computational: rulesets, compilers, and formal proof assistants return ground-truth feedback in seconds [\hyperlink{ref8}{8}, \hyperlink{ref13}{13}, \hyperlink{ref14}{14}]. In biomedicine, systems such as AI co-scientist [\hyperlink{ref1}{1}] and Virtual Lab [\hyperlink{ref2}{2}] have shown that \textit{in silico} debate, tournament ranking, and tool execution can effectively surface high-value hypotheses and candidate designs from literature priors and computational models. In molecular modalities where computational predictors are still maturing, however, pairing this \textit{in silico} reasoning with rapid physical measurement offers a way to ground agentic search directly in empirical data. Computational predictors sit on a spectrum of reliability dictated by their training data and cannot replace wet-lab validation for any modality [\hyperlink{ref17}{17}, \hyperlink{ref29}{29}]: even in protein engineering, where structure-based design succeeds on canonical rigid interfaces [\hyperlink{ref15}{15}, \hyperlink{ref16}{16}, \hyperlink{ref17}{17}], predictors struggle on flexible antibody loops and out-of-distribution targets [\hyperlink{ref29}{29}, \hyperlink{ref30}{30}], and models for nucleic acids and aptamers remain poorly calibrated for affinity and specificity due to sparse structural data [\hyperlink{ref18}{18}, \hyperlink{ref34}{34}]. Demonstrating closed-loop learning requires a design space where computational predictions alone fall short, paired with an experimental assay that is fast enough for multi-round iteration and returns quantitative, multiplexed readouts across both binders and non-binders.

Here we establish a closed-loop framework for agentic biomolecular discovery using single-stranded DNA aptamers. While earlier aptamer studies trained supervised or generative models on large SELEX datasets or coupled microarray screening with genetic algorithms and Bayesian optimization [\hyperlink{ref31}{31}, \hyperlink{ref32}{32}, \hyperlink{ref33}{33}, \hyperlink{ref36}{36}, \hyperlink{ref37}{37}], we ask whether LLM-based reasoning agents can design \textit{de novo} binders without prior target-specific training data, learn iteratively from pooled measurements, and compress the time and resources required by conventional selection [\hyperlink{ref19}{19}, \hyperlink{ref20}{20}]. Across three closed-loop design--test--refine rounds against $\alpha$-thrombin and myoglobin ($\sim$1,000 candidates per round), agentic pipelines improved statistically significant, bead-corrected on-target hit rates among newly designed candidates from 0.0\% (0/167) to 26.0\% (54/208) to 42.1\% (131/311) on $\alpha$-thrombin (37.1\% [106/286] across strictly \textit{de novo} candidates free of known literature motifs) and from 10.2\% (20/196) to 3.7\% (9/244) to 33.5\% (184/549) on myoglobin. Rather than screening $10^{13}\text{--}10^{15}$ random molecules by blind selection as in classical SELEX, this approach tests targeted pools of $\sim$1,000 designed sequences per round to return quantitative positive and negative feedback on every hypothesis. Orthogonal single-clone bead-based fluorescence titrations confirmed that pooled sequencing enrichment scales with physical binding ($r = 0.90$), yielding preliminary nanomolar affinities (apparent $K_D = 24.6\text{--}133\text{ nM}$) for agent-stabilized literature motifs and micromolar binding onset ($K_D > 1\ \mu\text{M}$) for strictly \textit{de novo} scaffolds. When multi-target counter-screening exposed a collapse in specificity driven by promiscuous binding motifs, the agents diagnosed the failure mode from raw sequencing data and pivoted to alternative structural families to restore target selectivity. These results establish the feasibility of closed-loop, agent-guided biomolecular discovery across multi-round experimental campaigns.

\section*{Results}

\subsection*{R.1 Aptamers are the ideal sandbox for lab-in-the-loop molecular design}

To test whether AI agents can learn iteratively from wet-lab data, we built a closed-loop pipeline around single-stranded DNA (ssDNA) aptamers (\textbf{Fig. 1}). In each cycle, an agentic design system equipped with literature search, target-structure lookup, biophysical folding tools, and a Python environment for writing code and running scientific libraries analyzes sequencing data from prior rounds and proposes a pool of $\sim$1,000 candidate sequences (\textbf{Fig. 1B}). Those candidates are synthesized as oligonucleotide pools, screened against immobilized target proteins by magnetic-bead pulldown, and read out by next-generation sequencing (NGS), returning quantitative enrichment profiles to the agents for the next round of design (\textbf{Fig. 1B, C}).

Aptamers are well suited for closed-loop molecular design because the nucleotide sequence \textit{is} the functional molecule (\textbf{Fig. 1A}). In protein and antibody campaigns, moving from a candidate sequence to a binding readout requires gene synthesis, cloning, transformation, cell culture and expression, cell lysis, column purification, and the binding assay itself. This seven-step cascade takes 3--6 weeks per round and introduces dropouts across expression, folding, solubility, and purification between what the model designs and what gets tested (\textbf{Fig. 1A}). Conventional SELEX, the traditional method for aptamer discovery, avoids cloning but relies on blind physical enrichment from random libraries ($10^{13}\text{--}10^{15}$ molecules) across 8--15 six-step selection rounds of binding, PCR amplification, and $\lambda$-exonuclease single-strand regeneration before sequencing readout [\hyperlink{ref19}{19}, \hyperlink{ref20}{20}], yielding minimal information on why individual sequences succeed or fail (\textbf{Fig. 1A}). By contrast, an agentic pipeline can evaluate a wide range of structural ideas and sequence variants \textit{in silico} before committing to a focused synthesis pool of $\sim$1,000 candidates. Our workflow pairs pooled oligo synthesis with deep sequencing of both the pre-selection input library and triplicate target and bare-bead pulldowns, compressing each wet-lab cycle from candidate to readout to three bench steps completed in $\sim$10 days (including oligo pool synthesis and delivery) at a cost of $\sim$\$3 per candidate (pooled oligo synthesis and reagent cost), decoupling synthesis bias from binding and returning quantitative positive and negative labels for every designed sequence in the pool (\textbf{Fig. 1A, C}).

\begin{figure}[htbp]
\centering
\includegraphics[width=\textwidth,height=0.66\textheight,keepaspectratio]{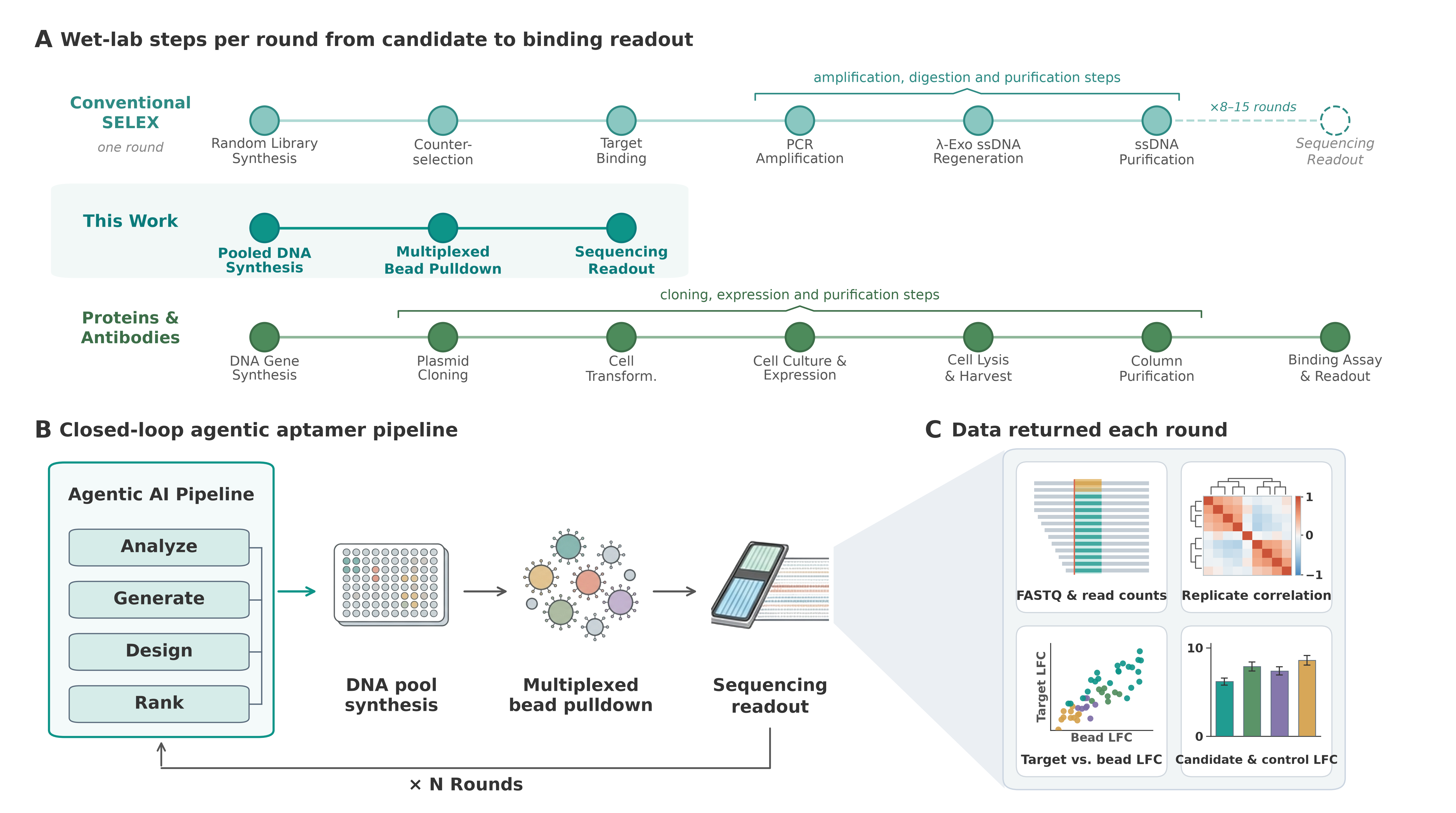}
\caption*{\textbf{Figure 1 | Aptamers as a fast, high-throughput sandbox for closed-loop agentic molecular discovery.} \textbf{(A)} Wet-lab steps per round from candidate to binding readout across modalities. Candidates designed by AI agents proceed directly from pooled DNA array synthesis to a multiplexed magnetic-bead pulldown and sequencing readout in 3 bench steps ($\sim$10 days per round), as an aptamer's nucleotide sequence is itself the functional molecule. This direct readout bypasses the 6-step per-round selection, amplification, and $\lambda$-exonuclease ssDNA regeneration cycle of conventional SELEX (repeated over 8--15 rounds before sequencing readout) and the 7-step cloning, expression, purification, and binding-assay cascade of protein/antibody pipelines. \textbf{(B)} Closed-loop agentic discovery pipeline ($\times N$ rounds). In each cycle, the agentic AI pipeline (\textit{Analyze}, \textit{Generate}, \textit{Design}, \textit{Rank}, transitioning from AI Co-Scientist zero-shot ideation in Round 1 to a domain-specialized tool-equipped agent harness in Rounds 2--3) designs $\sim$1,000 aptamer candidates for pooled oligo synthesis, which are assayed via multiplexed bead pulldown and next-generation sequencing (NGS) to return quantitative feedback for the next design round. \textbf{(C)} Schematic of the representative data and diagnostic package after each round, comprising demultiplexed FASTQ sequences, per-sequence read count matrices, and sample metadata returned from the wet-lab screen alongside the analyst agent's independently derived DESeq2 differential enrichment profiles, replicate/cross-target correlation matrices, and target-versus-bead $\log_2$ fold-change (LFC) regression diagnostics used to refine subsequent designs.}
\end{figure}

The aptamer sandbox turns molecular discovery into an iterative reasoning problem well suited to modern AI agent harnesses, returning clean positive and negative labels for $\sim$1,000 designs every $\sim$10 days within a fixed experimental budget. We asked whether agents could operate in a closed loop with the bench to discover binders \textit{de novo} and systematically improve their designs across rounds.

\subsection*{R.2 Hit rate improves from 0 to 42\% with the lab in the loop, outperforming zero-shot predictors}

We evaluated the closed-loop pipeline across three design--test--refine rounds on two proteins of contrasting difficulty: human myoglobin, a compact eight-helix bundle lacking a natural nucleic-acid binding cleft (\textit{de novo} target), and human $\alpha$-thrombin, a serine protease with two electropositive grooves that naturally bind nucleic acids (positive-control benchmark; \textbf{Fig. 2A; Methods; Supplementary Table S3}).

Across rounds, statistically significant, bead-corrected binder hit rates among newly designed candidates rose sharply on both targets (\textbf{Fig. 2B}). On myoglobin, agents discovered \textit{de novo} binders in Round 1 at a 10.2\% hit rate (20/196), dipped to 3.7\% (9/244) in Round 2 as non-specific G-quadruplex matrix binders were subtracted by counter-screening (Section R.3), and surged 9.1-fold in Round 3 to 33.5\% (184/549, $p = 9.6 \times 10^{-24}$; 100\% strictly \textit{de novo}; \textbf{Fig. 2B}). On $\alpha$-thrombin, where uninsulated Round 1 designs failed due to primer--core hybridization (0.0\%, 0/167), hit rates climbed to 26.0\% in Round 2 (54/208) and 42.1\% in Round 3 (131/311; 37.1\% [106/286] across strictly \textit{de novo} candidates free of known thrombin motifs; \textbf{Fig. 2B; Supplementary Table S3}). All random-sequence negative controls remained at 0.0\% across every round.

Directly beneath the hit-rate curves, raw target-versus-input enrichment distributions shifted toward stronger binding with each cycle (\textbf{Fig. 2C}). Prior to bare-bead subtraction, the fraction of each pool enriched at least twofold ($\log_2\text{FC} \ge 1$) rose monotonically in every round---climbing from 6.5\% to 12.4\% to 29.9\% on myoglobin and from 0.3\% to 53.1\% to 80.7\% on $\alpha$-thrombin (\textbf{Fig. 2C})---confirming that the Round 2 dip in myoglobin hit rate (\textbf{Fig. 2B}) reflected stringency filtering of sticky sequences rather than a loss of library enrichment.

To test whether computational predictors could have replaced these wet-lab cycles, we benchmarked 12 open-license predictor metrics (spanning 9 tools) zero-shot across the Round 3 myoglobin library ($N = 1{,}441$; \textbf{Fig. 2D}). Target-conditioned 3D co-folding (\texttt{Boltz-\allowbreak{}2} iPTM and $\Delta\text{iPTM}$: $\text{ROC-AUC} = 0.499$ and $0.391$), 3D inverse-folding and structure models (\texttt{ESMFold2-\allowbreak{}NA}, \texttt{RhoDesign}, \texttt{RiboDiffusion}: $0.496\text{--}0.510$), nucleic-acid foundation models (\texttt{DNABERT-\allowbreak{}2}, \texttt{HyenaDNA}, \texttt{RNA-\allowbreak{}FM}: $0.364\text{--}0.523$), and G-quadruplex scorers (\texttt{G4Hunter} $0.366$, \texttt{ViennaRNA} $\Delta\Delta G_{\text{G4}}$ $0.476$) all performed near or below random chance ($0.50$). Only target-agnostic 2D DNA thermodynamic stability ($-\Delta G_{\text{MFE}}$, $\text{ROC-AUC} = 0.729$) and folding purity ($-\text{ED}$, $0.763$) from \texttt{ViennaRNA} [\hyperlink{ref22}{22}] showed meaningful signal (\textbf{Fig. 2D}): \texttt{ViennaRNA} filters out unstructured coils and primer-invaded misfolds, but has no knowledge of the protein target and therefore cannot substitute for empirical wet-lab selection.

\begin{figure}[htbp]
\centering
\includegraphics[width=\textwidth,height=0.66\textheight,keepaspectratio]{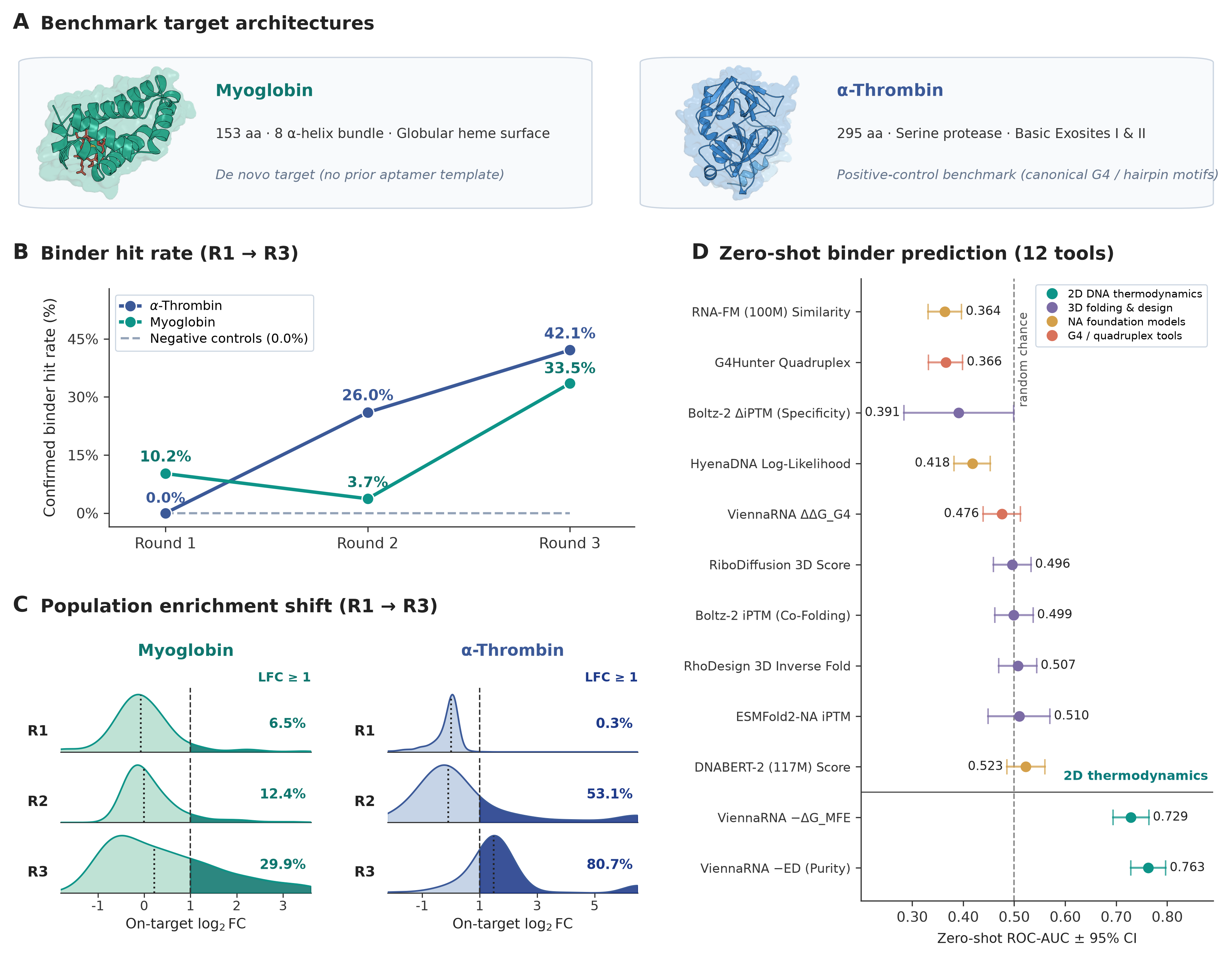}
\caption*{\textbf{Figure 2 | Closed-loop progression of binder hit rate, population enrichment, and zero-shot predictor benchmark across three rounds (R1 $\rightarrow$ R3).} \textbf{(A)} Surface structures and architecture of the benchmark targets: human myoglobin (left; PDB 3RGK, \textit{de novo} globular target) and human $\alpha$-thrombin (right; PDB 1PPB, positive-control benchmark with electropositive Exosites I and II). \textbf{(B)} Statistically significant, bead-corrected on-target binder hit rates across Rounds 1--3 for newly designed candidates on $\alpha$-thrombin ($0.0\% \rightarrow 26.0\% \rightarrow 42.1\%$) and myoglobin ($10.2\% \rightarrow 3.7\% \rightarrow 33.5\%$), alongside random-sequence negative controls ($0.0\%$, dashed line; full cohort and strictly \textit{de novo} breakdowns in \textbf{Supplementary Table S3}). \textbf{(C)} Kernel density distributions of raw target enrichment ($\log_2\text{FC}$ over input prior to bare-bead regression) across Rounds 1--3 for myoglobin (left) and $\alpha$-thrombin (right), stacked directly below \textbf{B}. Vertical dashed lines mark $\log_2\text{FC} = 1.0$, dotted lines mark population medians, and dark-shaded tails denote the fraction with $\log_2\text{FC} \ge 1.0$. \textbf{(D)} Zero-shot receiver operating characteristic area under the curve ($\text{ROC-AUC} \pm 95\%\text{ CI}$) across 12 open-license computational predictor metrics (spanning 9 tools) evaluated on the Round 3 myoglobin library ($N = 1{,}441$; 292 hits [20.3\%]). The vertical dashed line marks random chance ($\text{ROC-AUC} = 0.50$).}
\end{figure}

\subsection*{R.3 Agents diagnose G-quadruplex cross-reactivity and recover specificity via structural motif reallocation}

Multiplexed counter-screening explained the divergent Round 2 trajectories of the two targets (\textbf{Fig. 3A}). On $\alpha$-thrombin, whose electropositive Exosites I and II naturally recognize G-quadruplexes, clamping Round 1's weak folds lifted newly designed on-target hit rate from 0.0\% to 26.0\% and on-target specificity to 76.7\% in Round 2 (26.5\% on-target vs. 2.2\% off-target cross-reactivity in the counter-screened benchmark; \textbf{Fig. 3A}, right). On groove-less myoglobin, by contrast, expanding G-quadruplexes in Round 2 triggered a severe specificity collapse: 80.9\% (72/89) of binders in the counter-screened benchmark were cross-reactive (13.9\% off-target vs. 9.3\% on-target), crashing on-target specificity by 56 percentage points from 75.0\% in Round 1 to 19.1\% in Round 2 alongside the drop in \textit{de novo} hit rate from 10.2\% to 3.7\% (\textbf{Fig. 3A}, left; \textbf{Supplementary Table S3}).

From the Round 2 sequencing counts, the analyst agent diagnosed this failure mode without human input, finding that 97\% of cross-reactive binders were G-quadruplexes acting as non-specific polyanionic stickers (\textbf{Table 1}, Row 2; \textbf{Supplementary Figs. S1 and S2}). In Round 3, the agents therefore enforced bare-bead residual subtraction and reallocated both libraries across six structural motif families (\textbf{Fig. 3B}). On myoglobin, after keeping the library G4-heavy in Rounds 1--2 (58.2\%--60.7\% G4s), the agents cut unclamped G4s from 20.4\% to 1.3\% and pivoted 72.3\% of the Round 3 pool into non-G4 scaffolds (\textbf{Fig. 3B}, left). Whereas retained myoglobin G4s failed (0.0\%--2.4\% hit rate), non-G4 families drove the recovery---reaching an 80.4\% hit rate in H-type pseudoknots (37/46), 61.8\% in extended palindromic hairpins (47/76), and 45.9\% in bulge hairpins (67/146)---and accounted for 97.8\% (180/184) of Round 3 myoglobin hits with 0\% off-target cross-reactivity, restoring on-target specificity to 72.4\% and lifting \textit{de novo} hit rate 9.1-fold to 33.5\% (\textbf{Fig. 3A, B}, left). On $\alpha$-thrombin, after fixing Round 1 primer--core hybridization via $5'\text{--}3'$ duplex clamping in Round 2 (74.5\% G4s), the agents diversified Round 3 into 69.1\% non-G4 scaffolds and 30.9\% clamped G4s---achieving a 100.0\% hit rate in canonical TBA-stem G4 clamps (7/7), 57.5\% in extended palindromic hairpins (42/73), and 50.9\% in three-way junctions (29/57)---lifting on-target hit rate to 42.1\% and specificity to 98.5\% ($140\times$ selectivity over 0.3\% off-target cross-reactivity; \textbf{Fig. 3A, B}, right; \textbf{Supplementary Table S1}). All-by-all profiling in the Round 2--3 counter-screens, which also included human HER2 ectodomain as an off-target reference, surfaced a Round 2 myoglobin non-binder that enriched selectively on HER2. Carrying this seed into Round 3 alongside Watson--Crick-clamped chimeras yielded selective HER2 binders with no significant off-target enrichment on myoglobin or $\alpha$-thrombin (\textbf{Supplementary Fig. S3, Supplementary Note 1, Supplementary Table S2}).

\begin{figure}[htbp]
\centering
\includegraphics[width=\textwidth,height=0.66\textheight,keepaspectratio]{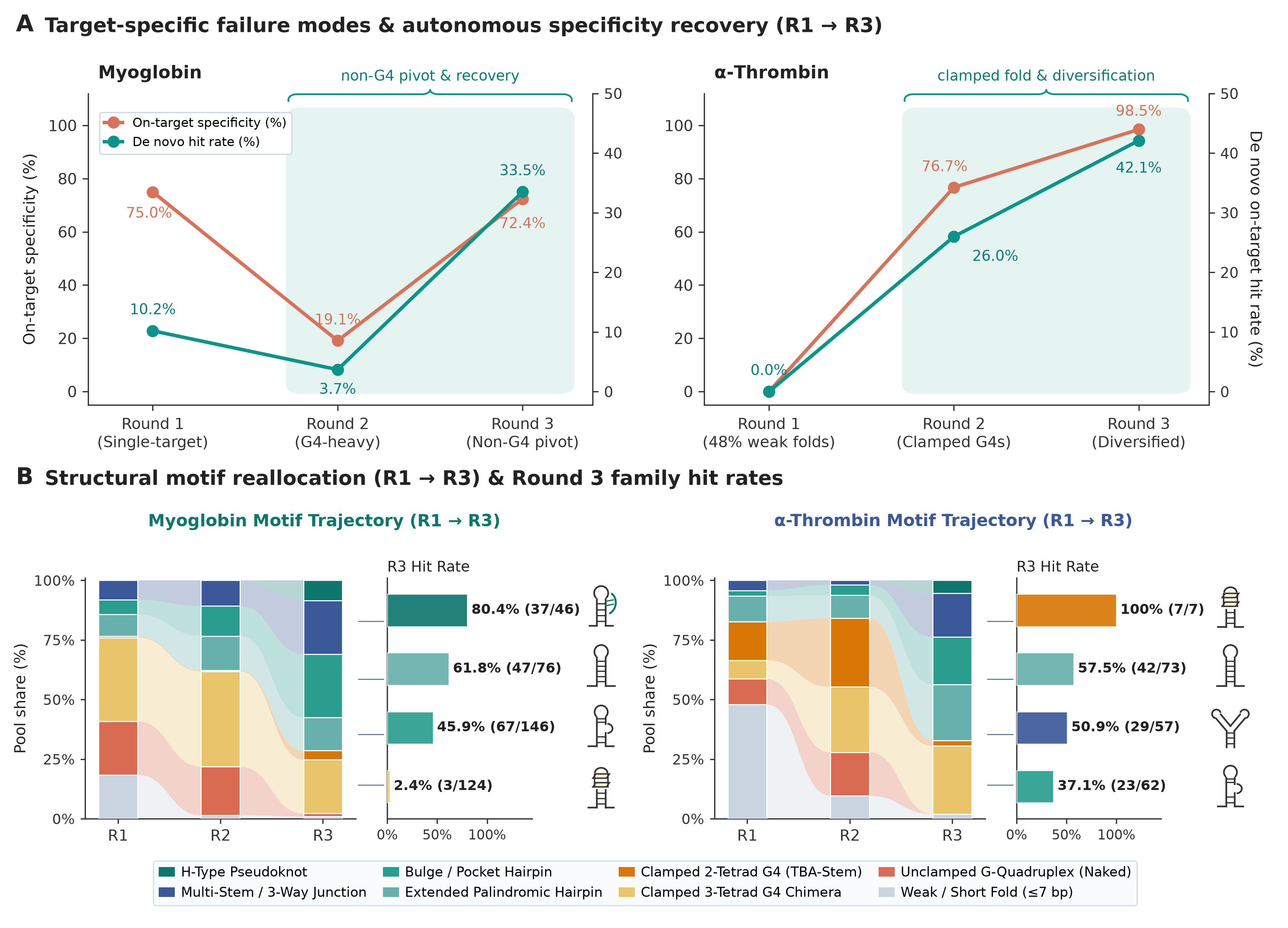}
\caption*{\textbf{Figure 3 | Autonomous diagnosis and recovery from G-quadruplex specificity collapse via target-matched structural motif reallocation (R1 $\rightarrow$ R3).} \textbf{(A)} Target-specific failure modes and autonomous specificity recovery across Rounds 1--3, showing on-target specificity (\%; coral curves, left axis; percentage of statistically significant binders in the counter-screened benchmark that bind exclusively to the intended target without off-target cross-reactivity) alongside newly designed on-target binder hit rate (\%; teal curves, right axis) for myoglobin (left; specificity $75.0\% \rightarrow 19.1\% \rightarrow 72.4\%$, \textit{de novo} hit rate $10.2\% \rightarrow 3.7\% \rightarrow 33.5\%$) and $\alpha$-thrombin (right; specificity $0.0\% \rightarrow 76.7\% \rightarrow 98.5\%$, newly designed hit rate $0.0\% \rightarrow 26.0\% \rightarrow 42.1\%$; full denominators in \textbf{Supplementary Table S3}). \textbf{(B)} Alluvial flow diagrams showing the reallocation of candidate structural motif families from Round 1 to Round 3 for myoglobin (left) and $\alpha$-thrombin (right), aligned column-for-column with \textbf{A} and paired with horizontal bar charts of Round 3 on-target hit rates ($n/N$) and predicted secondary-structure schematics for the top-performing families.}
\end{figure}

\subsection*{R.4 Agentic pipelines combine structural hypotheses with synthesis, assay, and statistical reasoning}

Beyond improving aggregate hit rates, inspecting the design rationales and campaign analyses across Rounds 1--3 reveals that the agents reasoned across the entire experimental stack (\textbf{Table 1}). Rather than acting as black-box sequence generators, the pipelines: (1) explored atypical \textit{de novo} structural topologies tailored to target surface geometry; (2) developed statistical counter-selection filters to separate specific binding from matrix noise and evaluated which computational scoring tools to trust; (3) proposed practical tweaks to wet-lab assay protocols, including liquid-handling automation and buffer optimization (logged as autonomous experimental recommendations while wet-lab buffer conditions were held constant across Rounds 1--3); and (4) connected \textit{in silico} sequence design to physical bottlenecks across solid-phase DNA synthesis, full-construct primer folding, and PCR amplification/sequencing recovery. The excerpts in \textbf{Table 1} illustrate reasoning traces and intermediate hypotheses generated by individual pipeline stages, documenting the prospective decision chain from Round 2 failure diagnosis to Round 3 library synthesis and validation. These are distinct from our post-hoc analyses (\textbf{Methods}) presented in \textbf{Figs. 2 and 3}, and were often refined or superseded in later steps. Progress likely came not from any single design rationale, but from iteratively pairing broad structural exploration with biophysical filtering and wet-lab feedback.

\begingroup
\footnotesize
\setlength{\tabcolsep}{5pt}
\renewcommand{\arraystretch}{1.15}
\vspace{8pt}\noindent{\small\bfseries \textbf{Table 1 | Examples of agentic scientific reasoning and documented decision chains across the closed-loop aptamer campaign.}\par}\vspace{-4pt}
\begin{longtable}{@{}>{\raggedright\arraybackslash}p{0.270\linewidth}>{\raggedright\arraybackslash}p{0.690\linewidth}@{}}
\toprule
\textbf{Reasoning Category \& Summary} & \textbf{Representative Agent Quotes \& Documented Decision Chain} \\
\midrule
\endhead
\bottomrule
\endfoot
\textbf{1. Explored atypical structural hypotheses} \par\smallskip Conducted \textit{de novo} explorations outside traditional single-loop hairpins and canonical motifs---reasoning directly about 3D scaffold geometry, inter-arm angles, directional bulges, and bipartite pockets to match target surface topology. & \textit{"Wide-hinge 44-nt three-way junction scaffold featuring dual 2-nt flexible single-stranded junction spacers. The flexible hinge geometry allows an expanded inter-arm angle (85--110$^\circ$) to wrap around the globular curvature of myoglobin without steric clash."} \par\smallskip \textit{"Dual-asymmetric internal loop (1-nt 5' vs. 4-nt 3' bulge) with an 8-bp GC basal clamp, creating an acute directional bend that positions the CAA apical loop and bulge into the myoglobin heme cleft."} \par\smallskip \textit{"13-bp duplex... project[s] the unstacked purine core (\texttt{TGAAAT}, positions 14--19) into solvent. The exocyclic faces of A16 and A17 are positioned to establish CH--$\pi$ and edge-to-face $\pi$--$\pi$ contacts against the exposed aromatic ring of Phe43... while the phosphodiester backbone at the T13:A20 neck forms electrostatic salt bridges with wild-type Lys45 (Arg45 in PDB 3RGK)."} \par\smallskip \textit{"Two interleaved stem-loops (H-type pseudoknot) creating a rigid, compact 3D scaffold with defined bipartite helical binding cavities."} \\
\addlinespace[4pt]
\arrayrulecolor{gray!35}\midrule[0.3pt]\arrayrulecolor{black}
\textbf{2. Leveraged statistical analysis and physics tools to guide design (Documented Round 2 $\rightarrow$ Round 3 Decision Chain)} \par\smallskip Built statistical counter-selection filters from raw counts to separate target-specific binding from magnetic-bead matrix noise, diagnosed G-quadruplex promiscuity and 3D co-folding failure, and shifted the synthesis portfolio toward non-G4 scaffolds prior to Round 3 screening. & \textbf{Observed Failure [Round 2 Screen] $\rightarrow$ Autonomous Diagnosis [Between-Round Analysis]}: In Round 2, myoglobin bead-corrected on-target hit rate dropped from \texttt{10.2\%} (\texttt{20/\allowbreak{}196}) in Round 1 to \texttt{3.7\%} (\texttt{9/\allowbreak{}244}) among newly designed myoglobin candidates (\texttt{9.3\%} [17/183] in the counter-screened benchmark including \texttt{8/\allowbreak{}16} carried-over R1 positive spikes; \texttt{9.0\%} [88/977] non-spike across the full pool; \texttt{80.9\%} [72/89] of binders cross-reactive, with \texttt{13.9\%} [72/517] off-target binding across all non-myoglobin designs). From raw count matrices, the analyst agent found that \textit{"The $R^2$ of 0.635 means 63.5\% of total enrichment variance is explained by non-specific bead affinity... Bead regression removes 85\% of false positives"} (in its initial QC-filtered analysis) and diagnosed that \textit{"thrombin-designed G-quads bind myoglobin at 18.7\% yield --- double the 9.3\% yield of myoglobin-designed sequences... NONE of the top-10 iPTM sequences overlap with actual binders... Using Boltz2 iptm as a design filter is actively harmful --- it selects for non-specificity."} \par\smallskip \textbf{Catching when linear regression breaks down [Between-Round Analysis]}: \textit{"Bead-binding regression is NOT appropriate for this [Thrombin] experiment --- not merely because $R^2$ is low (0.198), but because the regression slope is negative ($-$0.263), which causes regression to inflate false positives rather than remove them... Flag a binder if LFC\_beads / LFC\_thrombin > 0.5."} \par\smallskip \textbf{Proposed Change [Round 3 Strategy] $\rightarrow$ Implemented in Synthesis Order [Round 3 Pool]}: Capped pure/unclamped G-quadruplexes at $\le 15\%$ of the myoglobin design allocation (reducing unclamped naked G4s to \texttt{1.3\%} [7/549]), shifted \texttt{72.3\%} of the portfolio into non-G4 hairpins, pseudoknots, and three-way junctions (plus \texttt{26.4\%} clamped stem--G4 chimeras), calibrated \texttt{ViennaRNA} 2D folding (\textit{"Replace [3D co-folding] with ViennaRNA local folding in buffer conditions... actual binders are less stable [median MFE $-$5.1 vs $-$9.3 kcal/mol, p = 0.0007]"}), and embedded prospective \texttt{Boltz-\allowbreak{}2} calibration bands (\texttt{Tiers A-\allowbreak{}-\allowbreak{}D}) while humans held the wet-lab buffer constant. \par\smallskip \textbf{Subsequent Result [Round 3 Screen]}: Myoglobin on-target hit rate among newly designed candidates surged $9.1\times$ from \texttt{3.7\%} (\texttt{9/\allowbreak{}244}) to \texttt{33.5\%} (\texttt{184/\allowbreak{}549}; $3.6\times$ vs. the \texttt{9.3\%} [17/183] R2 benchmark and $3.3\times$ vs. R1 [\texttt{10.2\%}, 20/196], reaching \texttt{45.9\%-\allowbreak{}-\allowbreak{}80.4\%} across non-G4 hairpins and pseudoknots vs. \texttt{2.4\%} [3/124] for clamped 3-tetrad G4 chimeras and \texttt{0.0\%} [0/7] for unclamped G4s), specificity recovered from \texttt{19.1\%} to \texttt{72.4\%}, and prospective \texttt{Tier A} (\texttt{Boltz-\allowbreak{}2} $\text{iPTM} > 0.7$) yielded \texttt{0/\allowbreak{}25 (0.0\%)} statistically significant hits. \\
\addlinespace[4pt]
\arrayrulecolor{gray!35}\midrule[0.3pt]\arrayrulecolor{black}
\textbf{3. Suggested tweaks to wet-lab protocols} \par\smallskip Proposed concrete adjustments to liquid-handling automation, buffer ions, wash stringency, refolding, and control pooling \textit{(logged as autonomous protocol recommendations; wet-lab buffer held constant across Rounds 1--3)}. & \textit{"Implement automated magnetic bead washing protocols on robotic liquid handlers (e.g., Hamilton Star / Opentrons OT-2)... Standardize wash volume, agitation cycles, and temperature using automated liquid handlers. Automated wash execution directly prevents outlier replicates."} \par\smallskip \textit{"Our buffer is 1$\times$ PBS (4.5 mM $\mathrm{K}^+$); Thrombin TBA was selected in 25--100 mM KCl (our $\mathrm{K}^+$ is 5--20$\times$ too low)... Supplement binding buffer with 20 mM KCl."} \par\smallskip \textit{"Swap standard $\mathrm{K}^+$ and $\mathrm{Na}^+$ cations for $\mathrm{Li}^+$ during positive selection rounds to actively destabilize G-quadruplex folding, forcing enrichment of Watson-Crick-stabilized specific binders."} \par\smallskip \textit{"Increase stringency in magnetic bead washing (salt concentration, surfactant, and wash cycle count) and add 0.1 mg/mL salmon sperm DNA."} \par\smallskip \textit{"Switch aptamer refolding from bench snap-cool to slow ramp-down on a PCR cycler and increase binding incubation from 2 min on ice to 30 min at room temperature."} \par\smallskip \textit{"Evaluate positive controls strictly in isolated single-oligo lanes or spike at $\sim$1/100 so the candidate library remains measurable."} \\
\addlinespace[4pt]
\arrayrulecolor{gray!35}\midrule[0.3pt]\arrayrulecolor{black}
\textbf{4. Reasoned across design, synthesis, and experimental readout} \par\smallskip Connected \textit{in silico} sequence generation to physical constraints across solid-phase DNA synthesis, full-construct primer folding, and PCR amplification to prevent library dropouts and misfolding. & \textbf{Eliminating extreme palindromes to prevent library dropout}: \textit{"Of the wells that got zero reads, 96\% are self-complementary ($\Delta G$ of $-$35 to $-$47 kcal/mol, Fisher odds ratio $\approx$ 77, p = 5 $\times$ $10^{-165}$)... During base-by-base phosphoramidite chemistry at IDT, a strand that folds back on itself blocks the coupling step. In the regenerated R2 pool, high self-complementarity = 0 [dropout fell from 18.8\% to 1.8\%]."} \par\smallskip \textbf{85-nt full-construct primer invasion}: \textit{"Aptamer binding depends on the full oligo fold (85 nt), not just the 40-nt variable region. Fixed primer handles form competitive secondary structures ($\Delta G$ $\approx$ $-$6 to $-$12 kcal/mol) that destroy the aptamer's binding fold."} \par\smallskip \textbf{Spacer length \& palindromic penalty}: \textit{"Adding 7-nt TTTCGCG / CGCGTTT insulation spacers reduced sequencing read counts 2.7$\times$ to 5.9$\times$ by extending total oligo length and introducing palindromic CGCG motifs."} \par\smallskip \textbf{PCR stalling vs. synthesis dropout}: \textit{"Only 1.1\% of G4 sequences had zero input reads (synthesis failure); the remaining 74.0\% were present at low abundance due to PCR amplification bias. Without the valley QC filter, these sequences are no longer discarded --- DESeq2's negative binomial model accounts for them statistically."} \\
\addlinespace[4pt]
\arrayrulecolor{gray!35}\midrule[0.3pt]\arrayrulecolor{black}
\textbf{5. Evolved individual candidates and scaffolds across rounds} \par\smallskip Traced promising candidate lineages across selection rounds---diagnosing structural vulnerabilities in early seeds and re-engineering their flanking stems and recognition loops to improve affinity and specificity. & \textbf{Round 1 $\rightarrow$ Round 2 (Diagnosing primer invasion \& recovering active cores)}: Uninsulated R1 $\alpha$-thrombin cores failed (\texttt{0.0\%} hits) because fixed PCR primers hybridized with the binding motif; modeling full 85-nt constructs in R2 recovered the canonical 15-mer \texttt{TBA} core ($\log_2\text{FC} = +4.84$) and identified top G-quadruplex--hairpin hybrid seed \texttt{095e2a34c9a4} (embedding the 25-nt core of canonical \texttt{NU172}; $\log_2\text{FC} = +12.33$). \par\smallskip \textbf{Round 2 $\rightarrow$ Round 3 (Locking canonical \texttt{TBA} cores into Watson--Crick stem-clamps)}: Engineered 8--12 bp flanking duplex clamps around the R2 \texttt{TBA} core (\texttt{tba\_\allowbreak{}hairpin\_\allowbreak{}scaffold}) to prevent primer invasion and reduce bare-bead stickiness; in the subsequent Round 3 screen, all \texttt{22/\allowbreak{}22} (\texttt{100\%}) sequenced \texttt{tba\_\allowbreak{}hairpin\_\allowbreak{}scaffold} constructs were statistically significant, bead-corrected hits ($\log_2\text{FC} = +8.68\text{ to }+13.11$, comprising \texttt{5/\allowbreak{}5} short-stem \texttt{TBA-\allowbreak{}Stem} and \texttt{17/\allowbreak{}17} long-stem \texttt{Extended Palindromic Hairpin} constructs; \texttt{25/\allowbreak{}25} [\texttt{100\%}] across all R3 \texttt{TBA}/\texttt{NU172}-core designs), with representative 8-bp and 11-bp stem-clamped and loop-edited leads (\texttt{T02}--\texttt{T04}) validating in single-clone titrations at preliminary apparent $K_D = 47.4\text{--}133\text{ nM}$ (\textbf{Fig. 4}). \par\smallskip \textbf{Round 2 $\rightarrow$ Round 3 (Prospective mutational maturation of top hybrid seed)}: Generated systematic mutational variants (\texttt{R3\_\allowbreak{}095\_\allowbreak{}var1}--\texttt{var10}) around Round 2 champion \texttt{095e2a34c9a4} ($\log_2\text{FC} = +12.33$), introducing a 9-bp GC-rich terminal stem (\texttt{5'-\allowbreak{}TAACCAGCG...CGCTGGTTA-\allowbreak{}3'}) and a \texttt{C10$\rightarrow$A} strain-relieving bulge transition (\texttt{R3\_\allowbreak{}095\_\allowbreak{}var3}). In the subsequent Round 3 screen, \texttt{R3\_\allowbreak{}095\_\allowbreak{}var3} (\texttt{T01}) became the \#1 clone in the library (\texttt{44.55\%} of thrombin reads, \textbf{$\log_2\text{FC} = +16.27$}, $25.2\times$ specificity over bare beads) and validated in single-clone equilibrium titrations at a preliminary apparent \textbf{$K_D = 24.6 \pm 8.2\text{ nM}$} ($7.2\times$ fluorescence over background; \textbf{Fig. 4}). \\
\end{longtable}
\endgroup

\subsection*{R.5 Single-clone titrations confirm nanomolar thrombin binding}

We next tested whether enriched candidates bind their target independently of the pooled format, as pooled pulldown and sequencing readouts can be influenced by PCR amplification bias or competition effects. We individually synthesized eight 5'-fluorescein (5'-FAM)-labeled, full-length $\alpha$-thrombin candidates (\texttt{T01}--\texttt{T08}, 75--85 nt including their 45-nt constant primer flanks; \textbf{Supplementary Table S4}) spanning top-enriched, agent-engineered \texttt{NU172} and \texttt{TBA} architectures (\texttt{T01}--\texttt{T04}, $\log_2\text{FC} = +9.81\text{ to }+16.27$) and moderately enriched strictly \textit{de novo} scaffolds (\texttt{T05}--\texttt{T08}, $\log_2\text{FC} = +2.41\text{ to }+6.70$), and performed 5-point equilibrium titrations ($0\text{--}1{,}000\text{ nM}$) against bead-immobilized human $\alpha$-thrombin (\textbf{Fig. 4}). Six of the eight constructs (75\%) exhibited concentration-dependent binding above blank, and endpoint fluorescence at $1\ \mu\text{M}$ scaled quantitatively with Round 3 pooled enrichment across nearly four orders of magnitude (Pearson $r = 0.90$, $p = 0.0021$; Spearman $\rho = 0.88$, $p = 0.0038$; \textbf{Fig. 4B, C}), confirming that pooled $\log_2\text{FC}$ reflects genuine equilibrium binding rather than amplification artifacts. All four highly enriched candidates (\texttt{T01}--\texttt{T04}) yielded nanomolar binding curves (preliminary apparent $K_D = 24.6\text{--}133\text{ nM}$; \textbf{Fig. 4A}), whereas the moderately enriched strictly \textit{de novo} candidates (\texttt{T05}--\texttt{T08}) mapped to the micromolar regime expected for unoptimized, full-length \textit{de novo} hits prior to primer truncation (\textbf{Fig. 4B, C}). These fitted $K_D$ values should be considered preliminary estimates, as fluorescence signals did not reach a full saturation plateau ($B_{\max}$) within the $1\ \mu\text{M}$ concentration range tested.

Comparing clones across lineages directly validated the agents' structural engineering hypotheses (\textbf{Fig. 4A; Table 1, Row 5; Supplementary Tables S1 and S4}). The tightest binder, \texttt{T01} (\texttt{R3\_\allowbreak{}095\_\allowbreak{}var3}; $\log_2\text{FC} = +16.27$, $44.55\%$ of Round 3 thrombin reads), is an \texttt{NU172}-derived G4--duplex chimera with an agent-engineered \texttt{C10A} bulge transition and 9-bp GC-rich stem clamp that bound with a preliminary apparent $K_D = 24.6 \pm 8.2\text{ nM}$ ($\Delta\text{RFU} = +1{,}825$, $7.2\times$ over blank). Among the \texttt{TBA}-derived G4 constructs (\texttt{T02}--\texttt{T04}), extending the coaxial Watson--Crick stem clamp from 8 bp in \texttt{T03} ($\Delta G = -8.2\text{ kcal/mol}$) to 11 bp in \texttt{T02} ($\Delta G = -12.0\text{ kcal/mol}$) around the identical 15-nt core increased $1\ \mu\text{M}$ endpoint signal 2.6-fold ($\Delta\text{RFU} = +536$ vs. $+209$; preliminary $K_D = 47.4 \pm 14.6\text{ nM}$ vs. $132.7\text{ nM}$), while the central-loop variant \texttt{T04} (\texttt{G8A}, \texttt{TGT} $\rightarrow$ \texttt{TAT}), designed to reduce off-target enrichment in the pooled counter-screen, retained nanomolar binding to $\alpha$-thrombin (preliminary $K_D = 132.5 \pm 52.8\text{ nM}$). Within the strictly \textit{de novo} tier, both G4 scaffolds (\texttt{T05} and \texttt{T06}) showed reproducible binding onset at $1\ \mu\text{M}$ ($\Delta\text{RFU} = +118$ and $+58$; $K_D > 1\ \mu\text{M}$), whereas the two lower-enrichment non-G4 designs (\texttt{T07} and \texttt{T08}, $\log_2\text{FC} = +2.41\text{ and }+3.29$) remained near baseline ($\Delta\text{RFU} = +36$ and $+7$), consistent with the log-linear enrichment--affinity relationship (\textbf{Fig. 4C}).

\begin{figure}[htbp]
\centering
\includegraphics[width=\textwidth,height=0.66\textheight,keepaspectratio]{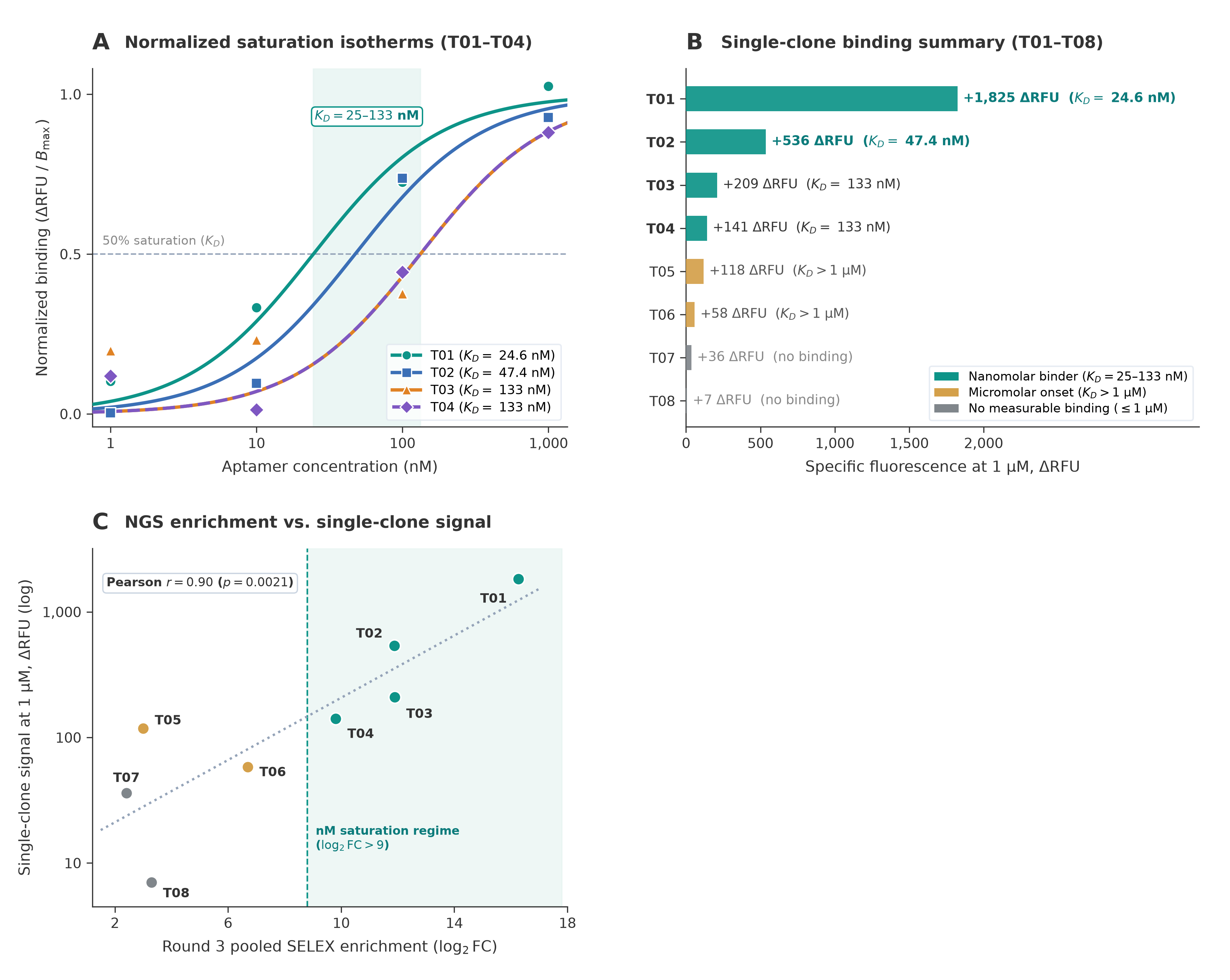}
\caption*{\textbf{Figure 4 | Orthogonal single-clone bead-binding validation of $\alpha$-thrombin aptamer candidates (\texttt{T01}--\texttt{T08}).} \textbf{(A)} Normalized equilibrium binding curves ($(\text{RFU} - y_0) / \Delta B_{\max}$) for the four nanomolar $\alpha$-thrombin binders (\texttt{T01}--\texttt{T04}) titrated across $1$, $10$, $100$, and $1{,}000\text{ nM}$ 5'-FAM-labeled full-length aptamer against biotinylated human $\alpha$-thrombin immobilized on streptavidin magnetic beads ($1\text{:}50$ bead dilution). Curves show non-linear least-squares 1:1 hyperbolic Langmuir fits ($\text{RFU}(C) = y_0 + \Delta B_{\max} \cdot C / (K_D + C)$); \texttt{T04} (preliminary $K_D = 132.5\text{ nM}$, dashed purple) is dashed to distinguish its curve from \texttt{T03} (preliminary $K_D = 132.7\text{ nM}$, solid orange). The horizontal dashed line indicates $50\%$ of fitted $\Delta B_{\max}$ (apparent $K_D$), and the shaded vertical band spans the preliminary nanomolar $K_D$ range ($24.6\text{--}133\text{ nM}$; treated as preliminary estimates as fluorescence did not reach an upper plateau within $1{,}000\text{ nM}$). \textbf{(B)} Blank-subtracted endpoint fluorescence at $1\ \mu\text{M}$ ($\Delta\text{RFU} = \text{RFU}_{1000\text{ nM}} - \text{RFU}_{0\text{ nM}}$) across all eight individually synthesized full-length candidates (\texttt{T01}--\texttt{T08}, 75--85 nt), color-coded by binding regime: nanomolar binders (\texttt{T01}--\texttt{T04}, teal), micromolar-onset \textit{de novo} binders (\texttt{T05}--\texttt{T06}, gold), and candidates below the $1\ \mu\text{M}$ titration ceiling (\texttt{T07}--\texttt{T08}, grey). \textbf{(C)} Quantitative concordance between Round 3 pooled sequencing enrichment ($\log_2\text{FC}$) and orthogonal single-clone fluorescence at $1\ \mu\text{M}$ ($\Delta\text{RFU}$, $\log_{10}$ scale) across \texttt{T01}--\texttt{T08} (Pearson $r = 0.90$, $p = 0.0021$). The vertical dashed line marks the enrichment threshold ($\log_2\text{FC} > 9$) separating candidates with preliminary nanomolar $K_D$ (\texttt{T01}--\texttt{T04}) from moderate-enrichment candidates in the micromolar regime (\texttt{T05}--\texttt{T08}). Full sequence details, raw RFU titrations, and curve-fitting parameters under both fixed-blank and free-baseline models are provided in \textbf{Supplementary Table S4}.}
\end{figure}

\section*{Discussion}

Our results demonstrate the feasibility of executing multi-round biomolecular discovery in a closed loop between AI agents and wet-lab assays, showing how iterative experimental feedback enables recovery from initial design failures. Across all biomolecular modalities, computational structure and binding predictors vary widely in accuracy depending on their training data, and none can replace experimental validation in the lab [\hyperlink{ref17}{17}, \hyperlink{ref29}{29}]. While structure-based models perform well on rigid, well-represented protein interfaces [\hyperlink{ref15}{15}, \hyperlink{ref16}{16}, \hyperlink{ref17}{17}], their accuracy drops on out-of-distribution targets, flexible loops such as antibody CDRs [\hyperlink{ref29}{29}, \hyperlink{ref30}{30}], and nucleic-acid aptamers where solved complex structures and negative binding data are scarce [\hyperlink{ref18}{18}, \hyperlink{ref24}{24}, \hyperlink{ref34}{34}]. Most agentic studies to date have been limited to a single experimental pass because protein and antibody campaigns require multi-week cloning, expression, and purification cascades [\hyperlink{ref1}{1}, \hyperlink{ref2}{2}, \hyperlink{ref3}{3}, \hyperlink{ref4}{4}, \hyperlink{ref12}{12}]. Single-stranded DNA aptamers, chemically synthesized affinity reagents used in diagnostics, biosensors, and analytical assays [\hyperlink{ref19}{19}, \hyperlink{ref20}{20}, \hyperlink{ref24}{24}], are well suited as a model modality to establish and stress-test the lab-in-the-loop paradigm. Within three experimental cycles and $\sim$1,000 candidates per round, our pipelines lifted statistically significant, bead-corrected on-target hit rates among newly designed candidates from 0.0\% (0/167) to 26.0\% (54/208) in Round 2 and 42.1\% (131/311) in Round 3 on $\alpha$-thrombin (37.1\% across strictly \textit{de novo} candidates free of published thrombin cores) and from 10.2\% (20/196) in Round 1 to 3.7\% (9/244) in Round 2 and 33.5\% (184/549) in Round 3 on myoglobin, reaching 80\%--100\% hit rates within top-performing structural families and up to a $140\times$ ratio of on-target to off-target hit rates. Rather than relying on blind selection over $10^{13}\text{--}10^{15}$ random molecules as in classical SELEX [\hyperlink{ref19}{19}, \hyperlink{ref20}{20}], testing targeted pools of $\sim$1,000 designed sequences per round returned a complete, counter-screened sequence--activity dataset of both positives and negatives across all $\sim$4,400 tested candidates (\textbf{Supplementary Table S3}).

Whereas earlier computational aptamer methods trained supervised or generative models on large SELEX datasets or mutated sequences under a fixed evolutionary or Bayesian objective [\hyperlink{ref31}{31}, \hyperlink{ref32}{32}, \hyperlink{ref33}{33}, \hyperlink{ref36}{36}, \hyperlink{ref37}{37}], the agents' search space within our loop extended to both the analytical workflow and atypical structural topologies that human designers rarely prioritize (Section R.4). Acting like bench experimentalists, the agents derived a two-stage statistical filter (differential enrichment gated by linear bead-background regression) that removed 36--55\% of apparent volcano pre-hits across full pools (and up to 71--85\% of raw enriched G-quadruplexes; \textbf{Table 1; Supplementary Figs. S1 and S2; Supplementary Table S3}), modeled full primer-flanked constructs [\hyperlink{ref22}{22}] to diagnose primer--core hybridization, built pass/fail quality-control gates to separate assay failure from design failure, and revised their own hypotheses after a single round of contradicting wet-lab data. When early selection rounds rewarded G-quadruplexes, which dominate published aptamer databases, multiplexed counter-screening exposed that 97\% of cross-reactive sequences were G-quadruplexes acting as promiscuous polyanionic surface stickers [\hyperlink{ref24}{24}, \hyperlink{ref26}{26}]. This finding prompted the agents to pivot toward designs predicted to form three-way junctions and H-type pseudoknots (80.4\% hit rate on myoglobin; \textbf{Fig. 3B}). This ability to dial between literature-guided exploitation on tractable targets ($\alpha$-thrombin) and atypical structural exploration on recalcitrant targets (myoglobin) gives human scientists high-level directional control over the discovery campaign.

These dynamics explain why computational models cannot yet substitute for physical measurement in nucleic-acid discovery, and they highlight the hazard of active learning on confounded data. Three-dimensional co-folding and structure-based models (\texttt{Boltz-\allowbreak{}2} [\hyperlink{ref21}{21}], \texttt{ESMFold2-\allowbreak{}NA}, \texttt{RhoDesign}, \texttt{RiboDiffusion}) and nucleic-acid foundation models performed near random chance ($\text{ROC-AUC} = 0.36\text{--}0.52$; \textbf{Fig. 2D}), consistent with known limitations caused by scarce aptamer crystal structures and lab-to-lab irreproducibility in public training corpora [\hyperlink{ref18}{18}, \hyperlink{ref24}{24}, \hyperlink{ref25}{25}, \hyperlink{ref26}{26}, \hyperlink{ref34}{34}]. In fact, raw \texttt{Boltz-\allowbreak{}2} interface confidence scores anticorrelated with specific binding because the model rewarded compact G-quadruplex folds that adhered non-specifically to solid supports (\textbf{Fig. 2D}). While 2D thermodynamic folding (\texttt{ViennaRNA} [\hyperlink{ref22}{22}]) showed moderate enrichment discrimination ($\text{ROC-AUC} = 0.73\text{--}0.76$; \textbf{Fig. 2D}), it is a purely sequence-intrinsic, target-agnostic stability calculator: it verifies that a DNA sequence folds into a clean secondary structure rather than a random coil, but has no knowledge of the protein target and therefore cannot predict target affinity or specificity. When a supervised classifier was trained on early selection rounds prior to bead-background subtraction, it learned G-tract frequency as a positive feature and inverted its accuracy on subsequent pools. Active learning built on un-decoupled assay data will therefore converge on experimental artifacts rather than genuine affinity. Chemically synthesized DNA pools partially mitigate this risk once paired with counter-selection: unlike protein screens where 60--80\% of designs fail to express or fold in cells [\hyperlink{ref15}{15}, \hyperlink{ref23}{23}], sequencing the input library alongside triplicate pulldowns verifies that every synthesized candidate that fails to enrich genuinely failed to bind, providing clean negative and positive labels for downstream model training.

The specificity recovery in Round 3 resolved indiscriminate cross-binding to streptavidin beads and unrelated proteins, restoring myoglobin on-target specificity from 19.1\% to 72.4\% while lifting $\alpha$-thrombin specificity from 76.7\% to 98.5\% (\textbf{Fig. 3A}). More broadly, this progression illustrates a key advantage of iterative closed-loop discovery over one-shot screening: rather than forcing agents or assays to satisfy all design constraints simultaneously, each cycle can layer on a new selective pressure. The loop can optimize first for target affinity, then for counter-screened specificity, and in subsequent rounds for activity in complex biological matrices such as serum, cell lysate, or cellular uptake. Catching and correcting assay-level matrix artifacts is an essential first gate, although fine discrimination against close structural homologues (such as myoglobin versus hemoglobin, or $\alpha$-thrombin versus prothrombin) was not tested here and remains a next step for counter-selection campaigns. Similarly, after an initial screen across a broader protein panel spanning clinical relevance and surface charge, the three proteins profiled in our rounds spanned a clear biophysical difficulty spectrum for nucleic-acid recognition: $\alpha$-thrombin [\hyperlink{ref27}{27}], which presents two strongly electropositive polyanion-binding grooves ($\text{pI} \approx 7.0\text{--}7.6$) and four active literature controls; myoglobin, a compact neutral-to-weakly-basic 8-helix bundle ($17\text{ kDa}$, $\text{pI} \approx 7.2$) lacking a natural nucleic-acid cleft where zero of four published controls bound; and the human HER2 ectodomain introduced in Round 2 counter-screening, a large ($70\text{ kDa}$), heavily glycosylated, net-electronegative receptor ($\text{pI} \approx 5.6$) that electrostatically repels the polyanionic DNA backbone and where all 24 published or literature-seeded controls failed (\textbf{Supplementary Fig. S3, Supplementary Note 1}). Final Round 3 hit rates and peak enrichments scaled monotonically across this gradient---from $\alpha$-thrombin (42.1\% on-target hit rate [37.1\% strictly \textit{de novo}], peak $\log_2\text{FC} = +16.27$) to myoglobin (33.5\% hit rate, $+3.61\ \log_2\text{FC}$) to strictly selective HER2 binders (5.1\% hit rate, $+4.59\ \log_2\text{FC}$, $3.8$-fold gain over Round 2)---while matching the distinct agentic strategies required at each tier: construct insulation on $\alpha$-thrombin, a topology pivot to non-G4 pseudoknots and hairpins on myoglobin, and cross-target seed repurposing with Watson--Crick clamping on HER2. The agents themselves diagnosed why the published myoglobin controls were inactive in our standardized pulldown assay [\hyperlink{ref24}{24}, \hyperlink{ref25}{25}, \hyperlink{ref26}{26}], tracing the discrepancy to primer-flank hybridization, physiological $\text{Na}^+/\text{K}^+$ buffer balance, and bead immobilization versus solution-phase selection (\textbf{Table 1}, Rows 3 and 4). They then recommended adding exogenous heme cofactor, substituting $\text{K}^+$, or adjusting wash stringency (Section R.4). We held physiological buffer conditions constant across all rounds to evaluate sequence-level learning on a fixed benchmark, though future loops should allow agents to co-optimize sequences and assay protocols.

On targets with prior literature such as $\alpha$-thrombin, the agents leveraged literature search and model priors to retrieve known \texttt{TBA} and \texttt{NU172} cores and re-engineer their flanking stems to survive full-construct folding, lifting their hit rate from 0.0\% in Round 1 to 100\% in Round 3 (25/25; \textbf{Table 1}, Row 5; \textbf{Supplementary Table S3}). Separating these literature-guided designs from strictly \textit{de novo} candidates confirmed that \textit{de novo} discovery also succeeded independently: strictly \textit{de novo} $\alpha$-thrombin designs reached a 37.1\% hit rate (106/286), and myoglobin---where no published controls bound---yielded a 33.6\% hit rate (184/547), with all hits strictly \textit{de novo}. Orthogonal single-clone bead-fluorescence titrations on $\alpha$-thrombin (\texttt{T01}--\texttt{T08}) further verified that Round 3 pooled sequencing enrichment translates quantitatively into single-clone equilibrium binding ($r = 0.90$, $p = 0.0021$; \textbf{Fig. 4; Supplementary Table S4}), calibrating the pooled sequencing assay against physical binding: 6 of 8 individually synthesized full-length constructs showed concentration-dependent binding, with agent-engineered stem clamps and loop edits of known motifs yielding preliminary nanomolar apparent affinities ($K_D = 24.6\text{--}133\text{ nM}$, \texttt{T01}--\texttt{T04}), moderately enriched \textit{de novo} G4 scaffolds (\texttt{T05}--\texttt{T06}) exhibiting clear binding onset at the $1\ \mu\text{M}$ assay ceiling ($K_D > 1\ \mu\text{M}$) even prior to constant-primer truncation, and lower-enrichment non-G4 \textit{de novo} designs (\texttt{T07}--\texttt{T08}, $\log_2\text{FC} = +2.41\text{--}3.29$) remaining below the $1\ \mu\text{M}$ detection threshold. Whereas prior supervised and generative machine learning methods yielded single-digit nanomolar binders ($K_D = 1.5\text{ nM}$) by training deep neural networks on nearly a million physical screening measurements [\hyperlink{ref36}{36}, \hyperlink{ref37}{37}], our closed-loop agents operated without any prior target-specific training data. In this zero-library context, achieving micromolar binding onset ($K_D > 1\ \mu\text{M}$) for strictly \textit{de novo} G4 scaffolds (\texttt{T05}--\texttt{T06}) within only three rounds ($\sim$3,000 focal candidates) represents an efficient initial discovery seed, contrasting with the nanomolar apparent affinities ($K_D = 24.6\text{--}133\text{ nM}$) achieved when the agents were provided a known starting motif (\texttt{TBA}/\texttt{NU172}) for targeted local optimization. As practical study caveats, commercial licensing constraints restricted our computational evaluations to open-license tools (Boltz-2, ESMFold2-NA, ViennaRNA, and open foundation models), and our single-clone titrations were performed exclusively on $\alpha$-thrombin and capped at $1\ \mu\text{M}$ on full-length (75--85 nt) bead-immobilized constructs without reaching an empirical saturation plateau ($B_{\max}$), rendering the fitted $K_D$ values preliminary. Extending single-clone validation to truncated minimal cores, additional closed-loop affinity-maturation rounds, and higher-concentration or label-free kinetic assays---specifically bio-layer interferometry (BLI) and surface plasmon resonance (SPR) ($k_{\text{on}}, k_{\text{off}}, K_D$) across myoglobin, $\alpha$-thrombin, and HER2---will be essential to biophysically confirm strictly \textit{de novo} leads, enabling physical affinity to be fed directly back into the closed loop alongside pooled NGS enrichment and counter-target specificity.

Our findings establish that coupling general-purpose and domain-specialized reasoning agents [\hyperlink{ref1}{1}, \hyperlink{ref7}{7}, \hyperlink{ref8}{8}, \hyperlink{ref9}{9}] with fast, counter-screened physical feedback enables autonomous scientific problem-solving across the full design--test--analyze loop. Across the multi-round campaign, we paired Google's AI co-scientist [\hyperlink{ref1}{1}] for initial zero-shot ideation in Round 1 with a tool-equipped aptamer harness for data-driven iteration in Rounds 2--3 rather than running a controlled comparison between agent architectures. While our campaign establishes closed-loop feasibility, prospective trials with matched-budget no-feedback and non-reasoning control arms will be needed to isolate the specific contribution of agentic reasoning, and head-to-head benchmarking against evolutionary, Bayesian, and supervised machine learning baselines [\hyperlink{ref31}{31}, \hyperlink{ref32}{32}, \hyperlink{ref33}{33}, \hyperlink{ref36}{36}, \hyperlink{ref37}{37}] is an important direction for future work. As reasoning models and agent harnesses continue to advance rapidly, the decisive determinant of discovery success is the speed and fidelity of the empirical feedback loop. Our three-step wet-lab workflow can be mapped end-to-end onto robotic liquid handlers coupled directly to sequencing because DNA pools and selection buffers are stable at room temperature [\hyperlink{ref10}{10}, \hyperlink{ref11}{11}, \hyperlink{ref12}{12}, \hyperlink{ref38}{38}], allowing agents to co-optimize sequences, buffers, cofactors, and stringency in a self-driving loop. Beyond aptamers, the closed-loop principles demonstrated in this aptamer sandbox are modality-agnostic, setting the stage for extending multi-round agentic discovery to more complex and resource-intensive assays such as protein--protein binder design [\hyperlink{ref12}{12}, \hyperlink{ref15}{15}, \hyperlink{ref16}{16}].

\section*{Methods}

\subsection*{Closed-loop pipeline overview}

Aptamer discovery was performed using a closed-loop pipeline in which AI agents designed candidate sequences, candidates were synthesized and experimentally screened, and sequencing read count matrices and sample metadata were returned to the agents to inform the next round of design. The pipeline comprised five stages executed in series: (i) agentic candidate design, (ii) pooled oligonucleotide array synthesis, (iii) multiplexed bead-based affinity pulldown, (iv) next-generation sequencing, and (v) return of count matrices and sample metadata to the agentic pipeline for iterative analysis and redesign.

Three complete design-test-improve cycles were performed, each completing in approximately 10 days of bench turnaround (\textbf{Supplementary Table S3}). Agents were tasked with generating approximately 640--1,450 candidate aptamer constructs per synthesized pool: 642 in Round 1, 981 in Round 2, and 1,450 ordered constructs in Round 3 yielding 1,441 sequenced library entries (1,405 after deduplicating secondary controls, comprising 1,208 newly designed constructs---549 primary myoglobin candidates, 311 primary $\alpha$-thrombin candidates, 200 exploratory HER2 candidates, 88 mutational seed variants, and 60 calibration tier constructs---plus 197 longitudinal positive, literature, and negative controls; 2,937 unique sequences across the three focal pools). Including the supplementary single-target myoglobin expansion pool synthesized between Rounds 1 and 2 (\texttt{Round 1b}, 1,353 ordered / 1,341 sequenced constructs), the campaign totaled 4,405 sequenced library constructs (4,425 ordered; 4,279 unique variable-region sequences after deduplicating cross-pool controls; \textbf{Supplementary Table S3}). Between rounds, human researchers executed the wet-lab pulldown, PCR amplification, and sequencing workflow, mapped sequencing reads to the designed reference library, and returned the resulting per-sequence read count matrices and sample metadata to the agent pipeline. Within the data-driven cycles (Rounds 2--3), differential enrichment analysis, cross-target and bare-bead background regression, structural classification, failure-mode diagnosis, constraint derivation, and candidate scoring were executed independently by the agent pipeline from the raw count data without receiving human-derived enrichment tables or analytical hypotheses, while wet-lab assay conditions were held constant across Rounds 1--3 by the human operators.

\subsection*{Agentic design platforms}

Two agentic platforms contributed sequentially across the three design rounds, with experimental sequencing counts and top-performing seeds shared across rounds so that each stage built directly on prior wet-lab outcomes.

\textbf{AI Co-Scientist.} Google's AI co-scientist [\hyperlink{ref1}{1}], a general-purpose multi-agent system designed for scientific ideation, multi-agent debate, and literature-grounded hypothesis generation, generated the initial Round 1 candidate library ($N = 642$) from target protein sequence and structural context, establishing \textit{de novo} binder discovery from zero prior experimental data. Top-performing Round 1 binders were carried forward into subsequent rounds as exploitation seeds and longitudinal controls.

\textbf{Custom aptamer pipeline.} For Rounds 2--3, where iterative improvement required ingesting prior-round sequencing count matrices, running independent statistical enrichment analysis in Python, performing structural classification and full-construct thermodynamic folding, and generating candidates subject to programmatic design constraints, we deployed a domain-specialized four-stage pipeline implemented on a modular Python agent orchestration framework with direct access to the computational toolbox described below. The pipeline employed a four-stage sequential architecture: (1) an \textit{analyst} agent that received raw per-sequence read count matrices and sample metadata and independently performed differential enrichment analysis (\texttt{PyDESeq2}), bare-bead background regression, ratio gating, cross-target profiling, and dropout/failure-mode diagnostics without receiving human-derived analytical results or structural hypotheses; (2) a \textit{strategist} agent that translated the analyst's report into target-specific design strategies with explicit explore-versus-exploit allocation; (3) \textit{parallel designer} agents, each assigned a specific strategy, that iteratively generated, folded, and filtered candidate sequences using the computational tools; and (4) a \textit{scoring} stage, executed outside the agent loop on cloud GPUs, that applied computationally intensive 3D structure-prediction models against target and decoy proteins to annotate candidates and construct prospective calibration tiers (\texttt{Tiers A-\allowbreak{}-\allowbreak{}D}).

\textbf{Language models.} The large language models used in the agentic pipelines were Google Gemini Pro and Flash models [\hyperlink{ref28}{28}], accessed via the Google Gemini API.

\subsection*{Computational tools available to agents}

Agents in the custom aptamer pipeline were provided access exclusively to computational tools carrying verified commercially permissive licenses (MIT, Apache-2.0, BSD-3-Clause, or the ViennaRNA commercial-permissive license); tools carrying non-commercial, copyleft, or unspecified licenses (such as RoseTTAFoldNA, HADDOCK, HDOCK, Nucleotide Transformer [CC-BY-NC-SA-4.0], AptaSuite [GPL-3.0], AptaTrans, and GRAPE-LM) were excluded. The toolbox comprised 9 computational modeling tools (yielding the 12 predictor metrics benchmarked in \textbf{Fig. 2D}) plus a structured protein-lookup utility, spanning four categories:

\textbf{2D DNA thermodynamics and structural topology classification.} ViennaRNA v2.6.4 (ViennaRNA Package License; explicit permission granted for research, educational, and commercial use [\hyperlink{ref22}{22}]) was used to compute minimum free energy (MFE) secondary structures and partition-function ensemble defect (ED) using the Mathews 2004 DNA nearest-neighbor thermodynamic parameter set (\texttt{dna\_\allowbreak{}mathews2004.par}) with G-quadruplex folding enabled (note that ViennaRNA's built-in quadruplex energy term is parameterized from RNA quadruplex data, and inverse-folding outputs were converted from RNA to DNA via uracil-to-thymine substitution). The six structural families in \textbf{Fig. 3B} were assigned by post-processing the predicted MFE dot-bracket structure and primary sequence, as standard dynamic-programming secondary-structure prediction folds only nested base pairs: predicted G-quadruplexes were classified as clamped ($\ge 8$ paired Watson--Crick stem bases) or unclamped ($<8$ paired bases); H-type pseudoknots were identified by scanning unpaired hairpin loops ($\ge 4\text{ nt}$) for $\ge 5\text{-nt}$ Watson--Crick reverse-complementarity to an unpaired $5'$ or $3'$ single-stranded tail ($\ge 6\text{ nt}$); and non-G4 duplexes were classified by stem and bulge topology into three-way junctions ($\ge 3$ stem segments or $\ge 2$ hairpins), bulge/pocket hairpins (containing internal bulge loops), or extended palindromic hairpins.

\textbf{3D structure prediction and inverse design.} Boltz-2 (v0.4, checkpoint \texttt{boltz2\_\allowbreak{}conf.ckpt}; MIT License [\hyperlink{ref21}{21}]) was used for protein--aptamer co-folding, providing interface predicted template modeling score (iPTM), predicted local distance difference test (pLDDT), predicted template modeling score (pTM), target-versus-decoy specificity contrast ($\Delta\text{iPTM}$), and a composite confidence score. ESMFold2-NA (HuggingFace checkpoint \texttt{biohub/\allowbreak{}ESMFold2-\allowbreak{}hf}; MIT License) provided nucleic-acid complex structure predictions. Candidate ssDNA variable regions were converted via thymine-to-uracil ($\text{T} \rightarrow \text{U}$) substitution prior to evaluation with \texttt{RhoDesign} (GitHub release \texttt{ml4bio/\allowbreak{}RhoDesign}, pretrained weights \texttt{rhodesign\_\allowbreak{}pretrained.pt}; MIT License) and \texttt{RiboDiffusion} (GitHub release \texttt{ml4bio/\allowbreak{}RiboDiffusion}, checkpoint \texttt{ribodiffusion\_\allowbreak{}sde.pth}; MIT License), which are parameterized on 3D RNA structures: \texttt{RhoDesign} computed 3D inverse-folding sequence log-likelihoods conditioned on a 44-nt pseudoknot backbone template (PDB 1A60), and \texttt{RiboDiffusion} scored 3-mer generative log-likelihoods derived from backbone-conditioned inverse-folding profiles converted back to DNA ($\text{U} \rightarrow \text{T}$).

\textbf{Nucleic acid foundation models.} DNABERT-2 (117M parameters, HuggingFace checkpoint \texttt{zhihan1996/\allowbreak{}DNABERT-\allowbreak{}2-\allowbreak{}117M} / \texttt{MAGICS-\allowbreak{}LAB/\allowbreak{}DNABERT\_\allowbreak{}2}; Apache-2.0 License) provided DNA embedding norms and seed cosine similarity scores, HyenaDNA (HuggingFace checkpoint \texttt{LongSafari/\allowbreak{}hyenadna-\allowbreak{}medium-\allowbreak{}160k-\allowbreak{}seqlen-\allowbreak{}hf}, \texttt{HazyResearch/\allowbreak{}hyena-\allowbreak{}dna}; Apache-2.0 License) contributed zero-shot causal autoregressive token log-likelihoods, and RNA-FM (100M parameters, \texttt{ml4bio/\allowbreak{}RNA-\allowbreak{}FM} checkpoint \texttt{RNA-\allowbreak{}FM\_\allowbreak{}pretrained.pth}; MIT License) evaluated $\text{T} \rightarrow \text{U}$ converted candidate sequences to compute layer-12 embedding cosine similarity against the literature seed.

\textbf{G-quadruplex and protein information.} G4Hunter scored G-quadruplex propensity from primary sequence using a direct Python implementation of the published 25-nt sliding-window equation (Bedrat et al., \textit{Nucleic Acids Res.} \textbf{44}, 1746--1759, 2016; CC-BY 4.0). A protein lookup tool returned structured knowledge for each target from open Protein Data Bank entries, including molecular weight, fold classification, surface electrostatic maps, binding-site residue coordinates, and published aptamer sequences.

Computationally intensive structure-prediction tools (\texttt{Boltz-\allowbreak{}2}, \texttt{ESMFold2-\allowbreak{}NA}) were executed on NVIDIA A100 SXM4 (40 GB) GPUs hosted on Google Cloud, whereas CPU-only tools (\texttt{ViennaRNA}, \texttt{G4Hunter}, sequence analysis utilities) ran within the agent's local Python environment.

\subsection*{Target proteins}

Three target proteins spanning a gradient of surface electrostatics, glycosylation, and nucleic-acid recognition tractability were profiled across the campaign. Human alpha-thrombin (36.7 kDa, $\text{pI} \approx 7.0\text{--}7.6$; PDB 1PPB) is a serine protease presenting two strongly electropositive anion-binding exosites (Exosite I and Exosite II) that constitute natural grooves for nucleic acid recognition, including well-characterized G-quadruplex aptamers. Thrombin served as a positive-control benchmark with known aptamer solutions. Human myoglobin (17.2 kDa, $\text{pI} \approx 7.2$; structural modeling reference PDB 3RGK, an engineered K45R/C110A crystallization variant sharing $>98\%$ sequence identity and an identical eight-$\alpha$-helix holo-myoglobin fold with wild-type human myoglobin assayed in the wet lab) is a compact globular protein coordinating a heme prosthetic group. Its surface presents only localized cationic patches and lacks any natural nucleic acid binding groove, making it a substantially more challenging target for \textit{de novo} aptamer discovery. Human HER2 (ErbB2) extracellular domain (ECD; subdomains I--IV, 630 aa, $\sim$70 kDa, $\text{pI} \approx 5.6$; PDB 1N8Z), introduced in Round 2 as a multiplexed counter-screen and continued in Round 3, is a large, heavily glycosylated receptor tyrosine kinase ectodomain whose net-electronegative surface at physiological pH electrostatically repels the polyanionic DNA backbone and lacks a canonical nucleic-acid binding cleft, serving both as an off-target specificity control and as a high-difficulty test case (\textbf{Supplementary Fig. S3, Supplementary Note 1}).

Biotinylated human alpha-thrombin was obtained from Assay Pro (catalog no. PT62011BIO), biotinylated human myoglobin was obtained from Molecular Depot (catalog no. P2010022B), and biotinylated human HER2 extracellular domain was obtained from Sino Biological (catalog no. 10004-HCCH-B-100).

\subsection*{Aptamer library design and synthesis}

Each round of the pipeline produced a pool of candidate aptamer sequences with variable regions of 30--50 nucleotides (target length $\sim$40 nt), flanked by a fixed 23-nt 5' forward constant region (\texttt{5'-\allowbreak{}TTGACTAGTACATGACCACTTGA-\allowbreak{}3'}) and a fixed 22-nt 3' constant region (\texttt{5'-\allowbreak{}TTCTGTCGTCCAGTCTGATGTG-\allowbreak{}3'}, written $5' \rightarrow 3'$ on the synthesized sense ssDNA strand) required for PCR amplification and sequencing library preparation, yielding full constructs of approximately 75--95 nucleotides ($\sim$85 nt). Variable region lengths were selected by the designing agents based on the structural architectures they chose to explore.

Candidate pools were synthesized as oligonucleotide arrays by Integrated DNA Technologies (IDT oPools). Synthesis orders were placed as received from the agentic pipeline without manual curation of individual sequences.

\subsection*{Multiplexed bead pulldown assay}

Binding selection was performed across parallel reaction wells as previously described [\hyperlink{ref35}{35}] with minor modifications. Dynabeads M-280 Streptavidin (Thermo Fisher Scientific, catalog no. 11206D) were washed and prepared following the manufacturer's instructions using $50\text{ pmol}$ of biotinylated target protein conjugated to the beads. Across all selection rounds, for each reaction, $5\text{ pmol}$ of refolded ssDNA was combined with the protein-conjugated (or bare control) beads. Following incubation, unbound aptamers were washed away, and bound aptamers were immediately amplified directly off the beads. PCR products were purified using Mag-Bind TotalPure NGS beads (Omega Bio-tek, catalog no. M1378) as previously described [\hyperlink{ref35}{35}] and quantified by Qubit fluorometry and Agilent TapeStation. Samples were then indexed using custom NGS primers, purified and pooled as previously described [\hyperlink{ref35}{35}], and sequenced on an Illumina MiSeq instrument ($1 \times 52\text{ bp}$ single-end reads).

Each pool was assayed simultaneously in parallel reactions against three core bead conditions: (i) alpha-thrombin-coated beads, (ii) myoglobin-coated beads, and (iii) bare streptavidin beads carrying no target protein (matrix background control), with each target-coated bead condition also serving as the off-target protein counter-screen for candidates designed against the other target. In Rounds 2 and 3, human HER2 extracellular domain (ECD; subdomains I--IV, $\sim$70 kDa, PDB 1N8Z) was also included in the multiplexed panel as an additional off-target protein reference (\textbf{Supplementary Fig. S3, Supplementary Note 1}). The technical replicate structure per sequencing run was: Round 1, $n = 3$ target / $n = 3$ bead / $n = 3$ input replicates per protein; Round 1b, $n = 2$ target / $n = 2$ bead / $n = 3$ input; Round 2, $n = 3$ target and $n = 3$ bead for $\alpha$-thrombin and HER2, $n = 4$ target and $n = 4$ bead for myoglobin, and $n = 3$ input across all assays; Round 3, $n = 3$ target / $n = 3$ bead / $n = 3$ input per protein. This multiplexed design enabled discrimination between specific target binding, non-specific matrix adhesion, and cross-reactive promiscuity.

\subsection*{Controls and reproducibility}

\textbf{Literature aptamer controls.} Eight published aptamers with reported binding activity against the study targets (four targeting alpha-thrombin and four targeting myoglobin) were synthesized alongside the AI-designed candidates and tested in the identical assay format. Of these, thrombin-binding aptamers TBA ($\log_2$ fold-change [LFC] $+8.22$ in Round 3), HD22 ($+3.03$), RE31 ($+3.87$), and NU172 ($+4.15$) showed strong enrichment on thrombin beads, verifying assay validity for that target. Zero of four myoglobin literature aptamers showed measurable enrichment on myoglobin beads.

\textbf{Random sequence controls.} Fifty-three random-sequence oligonucleotides of matched length and primer composition were included as negative controls. None (0/53) exceeded the enrichment threshold.

\textbf{Biological replicates.} To assess inter-operator reproducibility, the Round 3 candidate pool ($N = 1{,}441$ sequenced constructs; $1{,}405$ unique sequences) was independently screened by two operators in separate pulldown and sequencing runs. Continuous target-versus-input $\log_2\text{FC}$ enrichment scores correlated across operators ($r = 0.84\text{--}0.85$ across the full pool [$r = 0.840$ for myoglobin, $r = 0.835$ for $\alpha$-thrombin, and $r = 0.849$ for HER2]; $r = 0.920$ and $r = 0.964$ within the newly designed myoglobin [$n = 549$] and $\alpha$-thrombin [$n = 311$] candidate cohorts; and $r = 0.74\text{--}0.76$ on bare streptavidin beads and high-abundance subsets). Binary \texttt{DESeq2} enrichment pre-hit calls ($\log_2\text{FC} \ge 1.0$, $\text{FDR} < 0.05$) agreed between operators on 90.4\% of library constructs for myoglobin (Cohen's $\kappa = 0.740$; 345 vs. 357 pre-hits), 88.1\% for HER2 ($\kappa = 0.700$; 393 vs. 390 pre-hits), and 77.4\% for $\alpha$-thrombin (292 vs. 301 pre-hits; 84.9\% agreement [$\kappa = 0.439$] after bare-bead regression, with $r = 0.868$ between bead-subtracted residuals).

\subsection*{Next-generation sequencing, data return to agents, and uniform manuscript statistical evaluation}

Following bead pulldown and direct on-bead PCR amplification, indexed libraries across Rounds 1--3 were sequenced on an Illumina MiSeq platform ($1 \times 52\text{ bp}$ single-end reads). Sequencing reads were mapped to the designed reference library using a two-pass bioinformatics pipeline: Pass 1 assigned all exact ($d = 0$) variable-region matches (\texttt{\textasciitilde{}85-\allowbreak{}-\allowbreak{}90\%} of mapped reads), and Pass 2 applied $k$-mer-indexed Levenshtein fuzzy matching (edit distance $1 \le d \le 3$ nucleotides) exclusively to reads without an exact match, assigning each read to its unique minimum-distance reference construct and discarding reads tied at the minimum edit distance between two reference sequences. In Rounds 1 and 2, only 8.2\% (55/669) and 4.4\% (43/981) of unique variable regions had any library neighbor within edit distance $\le 3$; in Round 3 ($N = 1{,}405$), where single- and double-mutant maturation series ($n = 88$) were intentionally included, 28.2\% (396/1,405) had a neighbor within $d \le 3$, and exact-first priority matching with tie discarding prevented read bleed between point-mutant siblings.

The resulting raw per-sequence read count matrices and sample metadata were provided to the agentic pipeline as structured input for the subsequent design round. Within the closed loop between rounds, the agents wrote and executed their own analysis scripts (\texttt{PyDESeq2} differential enrichment, bare-bead regression, and ratio gating, as quoted in \textbf{Table 1}). \textbf{Separately from the agents' internal between-round scripts, all hit rates, specificity trajectories, volcano plots, and predictor benchmarks reported in this manuscript (Figures 2--3, Supplementary Table S3, and Supplementary Figures S1--S3) were computed by the authors using a standardized two-stage statistical pipeline applied across all rounds:}

\begin{enumerate}
  \item \textbf{Stage 1 (\texttt{DESeq2} differential enrichment over input)}: For each sequencing run and target condition $t$, raw replicate read counts for constructs passing the minimum detection filter ($\ge 10$ total reads across input and pulldown replicates) were normalized using \texttt{PyDESeq2} median-of-ratios size factors, modeled under Cox--Reid negative binomial dispersion estimation, and tested via Wald statistics with Benjamini--Hochberg false discovery rate ($\text{FDR}$) adjustment to obtain target-versus-input enrichment $y_{i,t} = \log_2\text{FC}_{i,\text{target vs. input}}$ ($\text{padj}_{i,t}$) and bare-bead-versus-input enrichment $x_i = \log_2\text{FC}_{i,\text{beads vs. input}}$ ($\text{padj}_{i,\text{beads}}$).
  \item \textbf{Stage 2 (Ordinary least-squares bare-bead background regression)}: The empirical slope $\beta_{1,t}$ relating $y_{i,t}$ to $x_i$ is systematically sub-unity ($\beta_{1,t} = 0.40\text{--}0.73$ across Rounds 2 and 3) because target-protein immobilization partially passivates exposed streptavidin matrix sites. We therefore fit an ordinary least-squares (OLS) linear model $\hat{y}_{i,t} = \beta_{0,t} + \beta_{1,t} x_i$ (calibrated in Round 3 on the $n = 102$ embedded negative, random, scrambled, and structural baseline controls: $\hat{y}_{i,\text{myo}} = -0.454 + 0.430 x_i$, $\hat{y}_{i,\text{thr}} = -0.960 - 0.067 x_i$, and $\hat{y}_{i,\text{her2}} = +0.130 + 0.681 x_i$) and computed each construct's bead-corrected residual $r_{i,t} = y_{i,t} - \hat{y}_{i,t}$. A construct was called a statistically significant, bead-corrected on-target hit (\texttt{is\_\allowbreak{}binder}, \textbf{Supplementary Table S3}) if it satisfied $\text{padj}_{i,t} < 0.05$, positive target enrichment over input ($\log_2\text{FC} \ge 1.0$ for volcano pre-hits), and a bead-corrected OLS residual $r_{i,t} > 1.00$ across all rounds (corresponding to $\ge 2$-fold enrichment above expected bare-bead background). For $\alpha$-thrombin, to prevent negative control intercepts or slopes from calling sequences that enriched on bare beads, Round 2 required bare-bead enrichment $x_i < 0.5$ (yielding an empirical minimum observed residual among called hits of $r_{i,\text{thr}} = +1.15$), and Round 3 required any construct that was also a significant bare-bead pre-hit ($\text{padj}_{i,\text{beads}} < 0.05$, $x_i > 1.0$) to exceed bare-bead enrichment by at least twofold directly ($y_{i,\text{thr}} - x_i > 1.0$; combined with the negative control intercept $\beta_{0,\text{thr}} = -0.960$ and $y_{i,\text{thr}} \ge 1.0$, this yielded an empirical minimum observed residual among called Round 3 $\alpha$-thrombin hits of $r_{i,\text{thr}} = +1.94$, as plotted in \textbf{Supplementary Fig. S1B, C} and noted in \textbf{Supplementary Table S3}). Comparing this two-stage OLS residual gate against a direct target-versus-bare-bead contrast ($\text{padj}_{i,t} < 0.05$, $y_{i,t} > 1.0$, and direct fold-change difference $\Delta\log_2\text{FC}_{t - \text{beads}} = y_{i,t} - x_i > 1.0$, which fixes $\beta_{1,t} = 1$) yielded 98.5\% binary hit-call agreement in Round 2 $\alpha$-thrombin ($\text{Jaccard} = 0.907$), 94.4\% in Round 2 myoglobin, 98.0\% in Round 3 $\alpha$-thrombin ($\text{Jaccard} = 0.903$), and 97.0\% in Round 3 HER2 ($\text{Jaccard} = 0.806$), while avoiding over-subtraction on myoglobin where bare-bead slope is $\beta_{1,\text{myo}} = 0.430$. Sensitivity analyses across relaxed ($\text{FDR} < 0.05$, $\log_2\text{FC} > 0.5$, $r_{i,t} > 0.5$: Round 2 $\rightarrow$ Round 3 myoglobin $6.6\%\ [16/244] \rightarrow 38.4\%\ [211/549]$ among new designs, $17.2\% \rightarrow 38.4\%$ across full/new benchmarks) and stringent ($\text{FDR} < 0.01$, $\log_2\text{FC} \ge 1.5$, $r_{i,t} > 1.5$: Round 2 $\rightarrow$ Round 3 myoglobin $2.5\%\ [6/244] \rightarrow 15.5\%\ [85/549]$ among new designs) thresholds confirmed that cross-round hit-rate gains are invariant to threshold choice. On-target specificity (\%) was defined as the percentage of all statistically significant binders in the counter-screened benchmark that bound exclusively to the intended target without significant off-target cross-binding.
\end{enumerate}

\subsection*{Single-clone fluorescence magnetic bead-binding assay and preliminary $K_D$ estimation}

To validate target binding, $\alpha$-thrombin aptamer candidates were tested as previously described with minor modifications [\hyperlink{ref35}{35}]. Briefly, $\alpha$-thrombin was immobilized on Dynabeads M-280 Streptavidin and refolded 5'-6-carboxyfluorescein (FAM) aptamers were incubated against the thrombin-coated beads across a logarithmic concentration series ($0, 1, 10, 100,\text{ and }1{,}000\text{ nM}$) for $1\text{ h}$ at room temperature with gentle agitation. Unbound oligonucleotides were removed by magnetic separation and single endpoint fluorescence was carried out using a plate reader. Preliminary apparent equilibrium dissociation constants ($K_D$) and fitted amplitude spans ($\Delta B_{\max}$) were estimated computationally as previously described [\hyperlink{ref35}{35}]. Fitted $K_D$ values should be considered preliminary estimates, as fluorescence signals did not reach an empirical saturation plateau ($B_{\max}$) within the $0\text{--}1{,}000\text{ nM}$ concentration range tested.

\section*{Data and code availability}

\textbf{Data availability.} Summary enrichment statistics, benchmark metrics, representative aptamer sequences, single-clone titration measurements, and curated agent reasoning traces supporting the findings of this study are provided within the manuscript, figures, \textbf{Table 1}, and \textbf{Supplementary Tables S1--S4}; complete de-identified sequence--activity and differential enrichment tables will be made available upon peer-reviewed publication. Atomic coordinates for the target proteins used in structural modeling are publicly available from the Protein Data Bank under accessions \href{https://www.rcsb.org/structure/1PPB}{1PPB} (human $\alpha$-thrombin), \href{https://www.rcsb.org/structure/3RGK}{3RGK} (human myoglobin), and \href{https://www.rcsb.org/structure/1N8Z}{1N8Z} (human HER2 extracellular domain).

\textbf{Code availability.} The closed-loop workflows described in this study are built on standard large language model APIs and multi-agent orchestration patterns rather than specialized proprietary algorithms; the four-stage architecture (analyst, strategist, parallel designers, and scoring), statistical filtering equations (\texttt{DESeq2} differential enrichment and ordinary least-squares bare-bead residual regression), and design constraints are detailed in the Methods and \textbf{Table 1} to allow independent reproduction on any tool-equipped agent platform. The \texttt{AI Co-\allowbreak{}Scientist} platform is described in detail in ref. [\hyperlink{ref1}{1}]. All computational folding, docking, and statistical tools invoked by the agents are publicly available open-source packages carrying commercially permissive licenses, including \texttt{ViennaRNA} v2.6 (\url{https://github.com/ViennaRNA/ViennaRNA}, ViennaRNA Commercial-Permissive License), \texttt{PyDESeq2} (\url{https://github.com/owkin/PyDESeq2}, MIT License), \texttt{Boltz-\allowbreak{}2} (\url{https://github.com/jwohlwend/boltz}, MIT License), \texttt{ESMFold2-\allowbreak{}NA} (\url{https://huggingface.co/biohub/ESMFold2-hf}, MIT License), \texttt{RhoDesign} (\url{https://github.com/ml4bio/RhoDesign}, MIT License), \texttt{RiboDiffusion} (\url{https://github.com/ml4bio/RiboDiffusion}, MIT License), \texttt{DNABERT-\allowbreak{}2} (\url{https://github.com/MAGICS-LAB/DNABERT_2}, Apache-2.0 License), \texttt{HyenaDNA} (\url{https://github.com/HazyResearch/hyena-dna}, Apache-2.0 License), and \texttt{RNA-\allowbreak{}FM} (\url{https://github.com/ml4bio/RNA-FM}, MIT License); \texttt{G4Hunter} scores were computed directly in Python from the published sliding-window equation (Bedrat et al., \textit{Nucleic Acids Res.} \textbf{44}, 1746--1759, 2016; CC-BY 4.0).

\section*{Biological safety and AI usage disclosure}

\textbf{Biological safety and biosecurity.} All target proteins screened in this study (human $\alpha$-thrombin, human myoglobin, and human HER2 ectodomain) are non-infectious, non-toxic human proteins studied under standard Biosafety Level 1 (BSL-1) laboratory conditions. All oligonucleotide pools (75--95 nt single-stranded DNA) were synthesized commercially by Integrated DNA Technologies (IDT) and underwent mandatory automated customer and sequence biosecurity screening in accordance with the International Gene Synthesis Consortium (IGSC) harmonized screening protocol and the U.S. Department of Health and Human Services (HHS) Framework for Nucleic Acid Synthesis Screening. No human subjects, clinical samples, or animal models were used in this study.

\textbf{Disclosure of generative AI usage.} In addition to serving as the core experimental subject of this study (closed-loop aptamer sequence generation, statistical enrichment analysis, failure diagnosis, and tool execution across Rounds 1--3, as detailed in Results and Methods), large language models (Google Gemini [\hyperlink{ref28}{28}]) were used by the authors to assist in manuscript editing, literature formatting, and figure layout generation. All experimental protocols, wet-lab assays, statistical definitions, and scientific conclusions were verified by the human authors, who take full responsibility and accountability for the integrity and content of this work.

\section*{Acknowledgements}

We thank Reuven Sayag, Matthias Heiler, Ronit Levavi Morad, and James Manyika for leadership and guidance; Dan Popovici, Philip Nelson, Michael Brenner, and John Platt for manuscript reviews and insightful feedback; Amanie Brik for overall contributions to the effort; and Artiom Myaskovsky and Alexander Daryin for additional technical contributions.

\section*{Author contributions}

Z.C., J.C., R.C., H.T., M.B., R.E., D.Z., P.S., A.H., Y.M., P.R., and A.P. conceived and supported the study. R.E., D.Z., P.S., P.R., and A.P. developed the custom agentic aptamer pipelines, computational tool integrations, and \texttt{AI Co-\allowbreak{}Scientist} workflows. Z.C., J.C., R.C., H.T., and M.B. designed and executed the multiplexed bead-based pulldown assays, next-generation sequencing library preparation, and bead-based fluorescence binding validation experiments. H.T., R.E., P.R., and A.P. performed the cross-round bioinformatics, statistical benchmarking, and agent log audits. A.H. and Y.M. provided research leadership and strategic guidance. P.R. and A.P. co-led the work. All authors contributed to writing, reviewing, and editing the manuscript.

\section*{Competing interests}

Z.C., J.C., R.C., H.T., M.B., R.E., D.Z., P.S., A.H., Y.M., P.R., and A.P. are employees or visiting faculty members of Google LLC, and some of these authors may own Alphabet Inc. stock as part of their standard compensation package. H.T. performed this work as a Visiting Faculty Researcher at Google Research. The authors declare no other competing interests.

\clearpage
\section*{Supplementary Figures and Tables}

\begin{figure}[htbp]
\centering
\includegraphics[width=\textwidth,height=0.66\textheight,keepaspectratio]{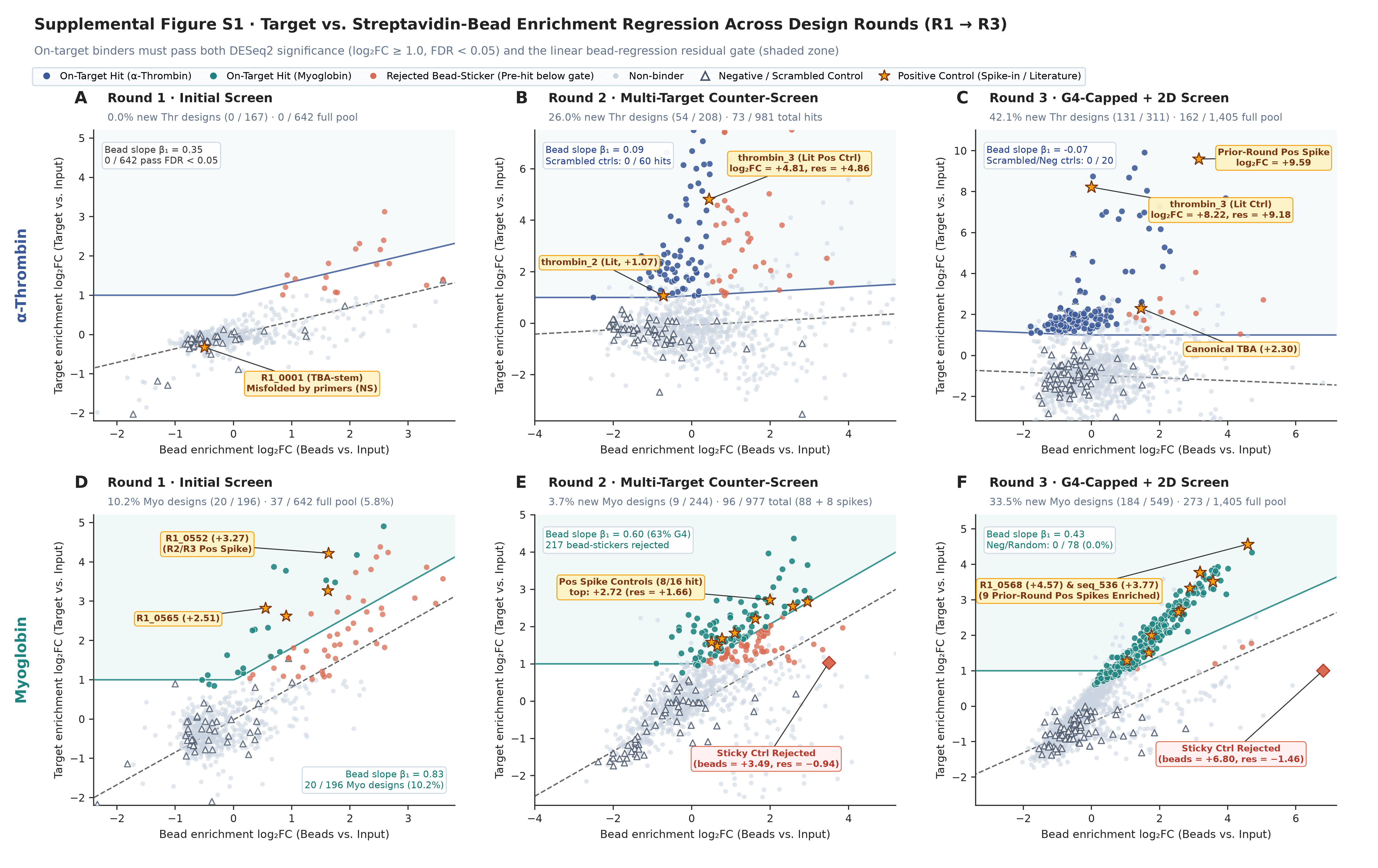}
\caption*{\textbf{Supplementary Figure S1 | Target-versus-streptavidin-bead enrichment regression and two-stage hit calling across design rounds (R1 $\rightarrow$ R3).} Target enrichment ($\log_2\text{FC}_{\text{target vs. input}}$, $y$-axis) plotted against bare streptavidin-bead enrichment ($\log_2\text{FC}_{\text{beads vs. input}}$, $x$-axis) for $\alpha$-thrombin (\textbf{A--C}) and myoglobin (\textbf{D--F}) across Round 1 ($N = 642$; \textbf{A}: $0/167$ [$0.0\%$] new $\alpha$-thrombin designs; \textbf{D}: $20/196$ [$10.2\%$] new myoglobin designs [$37/642 = 5.8\%$ full pool]), Round 2 ($N = 981$; \textbf{B}: $54/208$ [$26.0\%$] new $\alpha$-thrombin designs [$73/981$ total hits]; \textbf{E}: $9/244$ [$3.7\%$] new myoglobin designs [$96/977$ total hits including $8/16$ carried-over positive spikes]), and Round 3 ($N = 1{,}405$ deduplicated constructs [$1{,}441$ total library entries]; \textbf{C}: $131/311$ [$42.1\%$] new $\alpha$-thrombin designs [$162/1{,}405$ full pool]; \textbf{F}: $184/549$ [$33.5\%$] new myoglobin designs [$273/1{,}405$ full pool]; complete cohort breakdowns in \textbf{Supplementary Table S3}). Dashed lines indicate the ordinary least-squares bare-bead background trend (slope $\beta_1$); solid boundaries mark the specific binder zone requiring both $\log_2\text{FC}_{\text{target}} \ge 1.0$ (FDR < 0.05) and positive residual enrichment $r_{i,t} > 1.00$ above the bare-bead regression line (plus $x_i < 0.5$ in \textbf{B} and $y_{i,\text{thr}} - x_i > 1.0$ for bare-bead pre-hits in \textbf{C}). Points denote statistically significant $\alpha$-thrombin binders (blue circles), myoglobin binders (teal circles), rejected bead-stickers (coral circles, including 217 G4 designs in \textbf{E} with $\beta_1 = 0.60$), non-binders (grey dots), negative/scrambled/random controls (open triangles, $\triangle$), published and prior-round positive spike-in controls (gold stars, $\star$), and rejected non-specific sticky controls (red diamonds, $\diamond$ in \textbf{E--F}).}
\end{figure}

\begin{figure}[htbp]
\centering
\includegraphics[width=\textwidth,height=0.66\textheight,keepaspectratio]{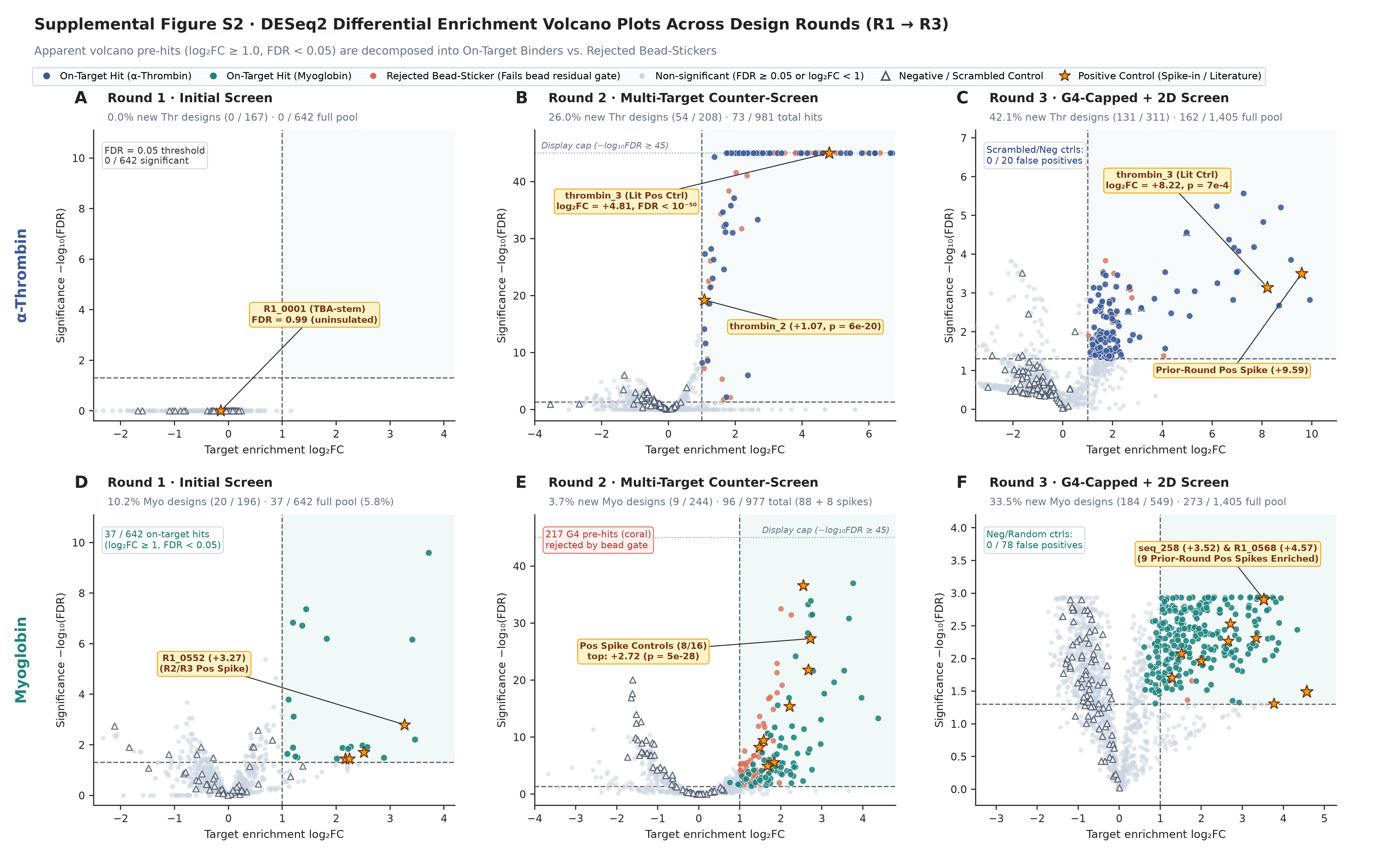}
\caption*{\textbf{Supplementary Figure S2 | DESeq2 differential enrichment volcano plots and control validation across design rounds (R1 $\rightarrow$ R3).} Statistical significance ($-\log_{10}\text{FDR}$, $y$-axis) versus target enrichment ($\log_2\text{FC}_{\text{target vs. input}}$, $x$-axis) for $\alpha$-thrombin (\textbf{A--C}) and myoglobin (\textbf{D--F}) across Rounds 1--3, with dashed lines marking the pre-hit cutoffs ($\log_2\text{FC} = 1.0$ and FDR = 0.05; upper-right quadrant). Apparent volcano pre-hits are decomposed into statistically significant on-target binders versus rejected bead-stickers eliminated by the bare-bead regression gate in \textbf{Supplementary Figure S1}. Stars ($\star$) mark published positive controls (uninsulated TBA stem in \textbf{A}; insulated thrombin controls TBA/\texttt{thrombin\_\allowbreak{}3} and \texttt{thrombin\_\allowbreak{}2} in \textbf{B--C}) and internal prior-round positive spike-in controls discovered in Round 1 (\textbf{D}) and carried forward into Rounds 2--3 (\textbf{E--F}). Open triangles ($\triangle$) mark random and scrambled negative controls (0\% false positives across all rounds). Dotted horizontal lines in \textbf{B} and \textbf{C} denote the display cap ($-\log_{10}\text{FDR} \ge 45$).}
\end{figure}

\begin{figure}[htbp]
\centering
\includegraphics[width=\textwidth,height=0.66\textheight,keepaspectratio]{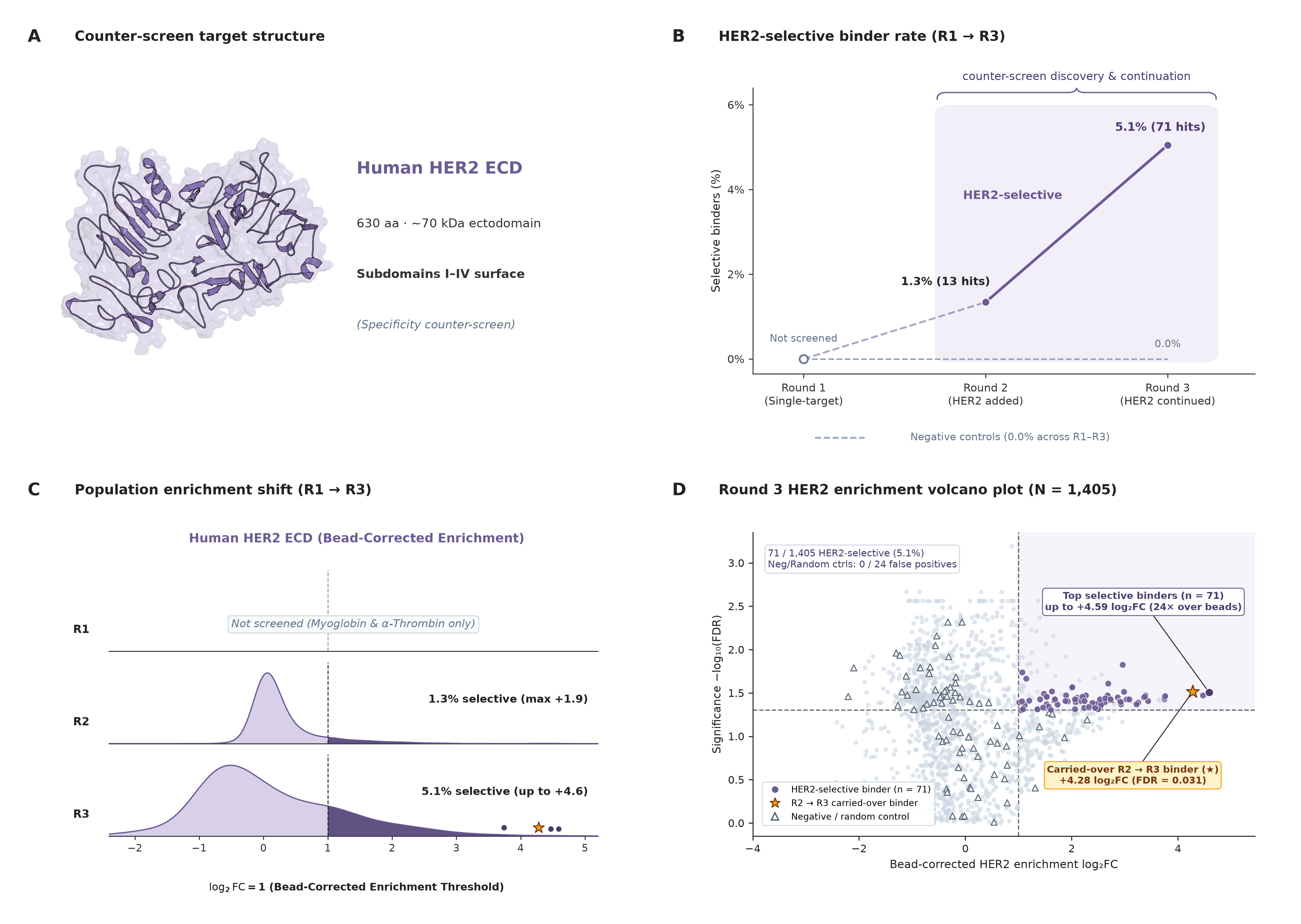}
\caption*{\textbf{Supplementary Figure S3 | Counter-screen discovery and continuation of HER2-selective binders across Rounds 1--3.} \textbf{(A)} Structure of the human HER2 extracellular domain (ECD; subdomains I--IV, $630\text{ aa}$, $\sim 70\text{ kDa}$, PDB 1N8Z), introduced in Round 2 as a specificity counter-screen and continued in the Round 3 multiplexed panel alongside myoglobin, $\alpha$-thrombin, and bare streptavidin beads. \textbf{(B)} Statistically significant HER2-selective binder rate across Round 1 (screened on myoglobin, $\alpha$-thrombin, and bare streptavidin beads; HER2 not yet screened), Round 2 ($N = 968$), and Round 3 ($N = 1{,}405$). Strictly HER2-selective binders ($\log_2\text{FC} \ge 1.0$, bead-corrected enrichment $\ge 1.0$, $\text{FDR} < 0.05$, with no significant enrichment on myoglobin or $\alpha$-thrombin) increased $3.8$-fold from 1.3\% (13 hits) in Round 2 to 5.1\% (71 hits) in Round 3, while negative controls remained at 0.0\% across all rounds (dashed line). \textbf{(C)} Kernel density distributions of bead-corrected HER2 enrichment ($\log_2\text{FC}$) across Rounds 1--3. The vertical dashed line marks the $\log_2\text{FC} = 1.0$ threshold; dark-shaded tails denote statistically significant selective binders ($\ge 1.0\ \log_2\text{FC}$), expanding from 1.3\% (max $+1.9\ \log_2\text{FC}$) in Round 2 to 5.1\% (reaching $+4.6\ \log_2\text{FC}$) in Round 3. Individual dots in the Round 3 right tail mark top selective binders ($\ge 3.5\ \log_2\text{FC}$), including the carried-over Round 2 $\rightarrow$ Round 3 binder ($\star$). \textbf{(D)} Round 3 volcano plot of bead-corrected HER2 enrichment ($\log_2\text{FC}$) versus statistical significance ($-\log_{10}\text{FDR}$) across all $N = 1{,}405$ candidates, highlighting the 71 HER2-selective binders (solid circles, up to $+4.59\ \log_2\text{FC}$ or $24\times$ over bare beads), the carried-over Round 2 $\rightarrow$ Round 3 binder (gold star $\star$, $+4.28\ \log_2\text{FC}$, $\text{FDR} = 0.031$), and negative/random controls (open triangles, $0/24$ false positives).}
\end{figure}

\subsection*{Supplementary Note 1 | Counter-screen discovery and continuation of HER2-selective binders across Rounds 2--3}

To evaluate target specificity and cross-reactivity across an expanded protein panel, human HER2 extracellular domain (ECD; subdomains I--IV, $\sim 70\text{ kDa}$) was introduced as an off-target reference in the Round 2 multiplexed counter-screen ($N = 968$) and continued in Round 3 ($N = 1{,}405$) alongside $\alpha$-thrombin, myoglobin, and bare streptavidin beads (\textbf{Supplementary Fig. S3A}).

While most G-quadruplexes in Round 2 behaved as non-specific matrix or polyanionic cross-binders, all-by-all profiling surfaced 13 strictly HER2-selective candidates (1.3\% of the pool; \textbf{Supplementary Fig. S3B}), including a failed myoglobin candidate (\texttt{bef5bce6520a}, $\star$) carrying a 6-bp antiparallel Watson--Crick stem (\texttt{5'-\allowbreak{}GAGCGG-\allowbreak{}3'} paired with \texttt{5'-\allowbreak{}CCGCTC-\allowbreak{}3'}) clamping a compact G-quadruplex core (\texttt{5'-\allowbreak{}AA-\allowbreak{}GAGCGG-\allowbreak{}GGTGGGTTTTCTTTGGGTTGGG-\allowbreak{}CCGCTC-\allowbreak{}AAAA-\allowbreak{}3'}) that showed no enrichment on myoglobin or bare beads ($\le -0.46\ \log_2\text{FC}$) but enriched selectively on HER2 ($+1.66\ \log_2\text{FC}$ raw, $+1.19\ \log_2\text{FC}$ bead-corrected, $\text{FDR} = 6.7 \times 10^{-5}$; \textbf{Supplementary Fig. S3C}). Carrying this seed into Round 3 alongside 6--8 bp Watson--Crick-clamped G-quadruplex chimeras shifted the population enrichment tail rightward (\textbf{Supplementary Fig. S3C}) and expanded HER2-selective binders $3.8$-fold to 5.1\% (71/1,405; \textbf{Supplementary Fig. S3B, D}). The carried-over Round 2 seed ($\star$) enriched again in Round 3 at $+4.28\ \log_2\text{FC}$ ($19\times$ over bare beads, $\text{FDR} = 0.031$), and newly designed 6--8 bp stem-clamped G-quadruplex chimeras reached $+3.22\text{ to }+4.59\ \log_2\text{FC}$ ($9\times\text{ to }24\times$ over bare beads) with no significant enrichment detected on myoglobin or $\alpha$-thrombin (\textbf{Supplementary Table S2}), while all negative controls remained at 0.0\% (0/24 false positives).

\begingroup
\scriptsize
\setlength{\tabcolsep}{3.5pt}
\renewcommand{\arraystretch}{1.15}
\vspace{8pt}\noindent{\small\bfseries Supplementary Table S1 | Representative aptamer motifs, sequence lineages, and structural hypotheses across Rounds 1--3\par}\vspace{-4pt}
\begin{longtable}{@{}>{\raggedright\arraybackslash}p{0.100\linewidth}>{\raggedright\arraybackslash}p{0.140\linewidth}>{\raggedright\arraybackslash}p{0.130\linewidth}>{\raggedright\arraybackslash}p{0.210\linewidth}>{\raggedright\arraybackslash}p{0.160\linewidth}>{\raggedright\arraybackslash}p{0.190\linewidth}@{}}
\toprule
\textbf{Target} & \textbf{Round \& Lineage} & \textbf{Structural Family} & \textbf{Variable Region / Conserved Motif (\texttt{5' $\rightarrow$ 3'})} & \textbf{Folding \& Enrichment Metrics} & \textbf{Structure--Activity Hypothesis} \\
\midrule
\endhead
\bottomrule
\endfoot
\textbf{$\alpha$-Thrombin} & Round 1 $\rightarrow$ Round 2 Primer-Insulation Fix (\texttt{T02}--\texttt{T04}) & Uninsulated vs. stem-clamped 2-tetrad G4 (\texttt{TBA} core) & Core: \texttt{GGTTGGTGTGGTTGG} (flanked by fixed primers in R1 vs. 6--11 bp duplex clamps in R2--R3) & R1 full-construct primer--core pairing $\Delta G = -7.8\text{ kcal/mol}$ (0.0\% hits) $\rightarrow$ R3 clamped stem 100\% hit rate (7/7 canonical 6--10 bp \texttt{TBA} clamps; 100\% [25/25] across all R3 clamped \texttt{TBA}/\texttt{NU172}-core constructs); preliminary single-clone $K_D = 47.4 \pm 14.6\text{ nM}$ (\texttt{T02}, 11-bp stem) vs. $132.7\text{ nM}$ (\texttt{T03}, 8-bp stem) and $132.5 \pm 52.8\text{ nM}$ (\texttt{T04}, \texttt{G8A} loop edit; \textbf{Fig. 4}) & Folding the full 85-nt primer-flanked construct in \texttt{ViennaRNA} indicated that fixed primers could hybridize with uninsulated cores; adding protective Watson--Crick stems restored predicted folding and nanomolar physical binding. \\
\addlinespace[4pt]
\arrayrulecolor{gray!35}\midrule[0.3pt]\arrayrulecolor{black}
\textbf{$\alpha$-Thrombin} & Round 2 G4--Hairpin Hybrid Seed (\texttt{NU172} core) & G-quadruplex--hairpin hybrid motif & Conserved 25-nt motif: \texttt{\seqsplit{GCCTAGGTTGGGTAGGGTGGTGGCG}} (25-nt core of canonical 26-nt \texttt{NU172}) & Shared across 17 of top 18 R2 binders ($\log_2\text{FC} > 11$) & Embeds the 25-nt G-quadruplex/duplex core of \texttt{NU172} within a flanking Watson--Crick stem clamp that insulates the motif from primer interference. \\
\addlinespace[4pt]
\arrayrulecolor{gray!35}\midrule[0.3pt]\arrayrulecolor{black}
\textbf{$\alpha$-Thrombin} & Round 3 Top Enriched Binder (\texttt{R3\_\allowbreak{}095\_\allowbreak{}var3} / \texttt{T01}) & Extended palindromic hairpin / clamped G4 hybrid & \texttt{\seqsplit{TAACCAGCGACTAGGTTGGGTAGGGTGGTGGCGCTGGTTA}} & $\log_2\text{FC} = +16.27$ (strongest binder of campaign, $44.55\%$ of R3 pool); preliminary single-clone $K_D = 24.6 \pm 8.2\text{ nM}$ ($\Delta\text{RFU} = +1{,}825$, $7.2\times$ over blank; \textbf{Fig. 4}); family hit rate 57.5\% (42/73) & Extending the Round 2 25-nt hybrid seed (\texttt{095e2a34c9a4}) with a 9-bp terminal stem (\texttt{TAACCAGCG...CGCTGGTTA}) and a \texttt{C10$\rightarrow$A} strain-relieving bulge maximized predicted folding purity, pooled enrichment, and single-clone nanomolar affinity. \\
\addlinespace[4pt]
\arrayrulecolor{gray!35}\midrule[0.3pt]\arrayrulecolor{black}
\textbf{Myoglobin} & Round 2 Basal-Stem vs. Loop Contrast & Stem-loop hairpin scaffold & Shared 14-bp basal stem: \texttt{GTCGAAGAGTTCAC} (paired with \texttt{GTGAACTCTTCGAC}) & 72/74 stem-matched designs failed; 2/74 enriched strongly ($\log_2\text{FC} = +2.49, +2.39$; $\Delta G_{\text{MFE}} < -10\text{ kcal/mol}$) & Comparing 74 sequences with identical basal stems indicated that stem stability alone is insufficient for binding and that enrichment depends on the 10--13 nt apical/internal loop sequence. \\
\addlinespace[4pt]
\arrayrulecolor{gray!35}\midrule[0.3pt]\arrayrulecolor{black}
\textbf{Myoglobin} & Round 3 Top Pseudoknot Lineage & H-type pseudoknot (\texttt{-\allowbreak{}CGCG-\allowbreak{}} tertiary lock) & \texttt{\seqsplit{GTCGAAGAGTTCACACGCGATGTGAACTCTTCGACATCGT}} & $\Delta G_{\text{MFE}} = -24.9\text{ kcal/mol}$; $\log_2\text{FC} = +3.61$; family hit rate 80.4\% (37/46) & Evolving the Round 2 stem-loop winner via constrained inverse folding into a predicted \texttt{-\allowbreak{}CGCG-\allowbreak{}} H-type pseudoknot stabilized loop presentation while eliminating G4 cross-reactivity. \\
\addlinespace[4pt]
\arrayrulecolor{gray!35}\midrule[0.3pt]\arrayrulecolor{black}
\textbf{Myoglobin} & Round 3 Non-G4 Hairpin Families & Extended palindromic \& bulge/pocket hairpins & Variable 8--12 bp stems interrupted by asymmetric internal bulges & Extended palindromic hit rate 61.8\% (47/76); bulge/pocket hairpin hit rate 45.9\% (67/146) & Replacing polyanionic G-quadruplexes (clamped 3-tetrad G4 chimeras: 2.4\% hit rate, 3/124; unclamped G4s: 0.0\% hit rate, 0/7) with non-G4 hairpins and 3WJs restored on-target specificity from 19.1\% to 72.4\%. \\
\end{longtable}
\endgroup

\begingroup
\scriptsize
\setlength{\tabcolsep}{2.8pt}
\renewcommand{\arraystretch}{1.15}
\vspace{8pt}\noindent{\small\bfseries Supplementary Table S2 | Top statistically significant HER2-selective binders in Round 3\par}\vspace{-4pt}
\begin{longtable}{@{}>{\raggedright\arraybackslash}p{0.150\linewidth}>{\raggedright\arraybackslash}p{0.125\linewidth}>{\raggedright\arraybackslash}p{0.185\linewidth}>{\raggedright\arraybackslash}p{0.115\linewidth}>{\raggedright\arraybackslash}p{0.060\linewidth}>{\raggedright\arraybackslash}p{0.070\linewidth}>{\raggedright\arraybackslash}p{0.060\linewidth}>{\raggedright\arraybackslash}p{0.155\linewidth}@{}}
\toprule
\textbf{Candidate ID} & \textbf{Round Origin \& Architecture} & \textbf{Variable Region Sequence (\texttt{5' $\rightarrow$ 3'}, 40--42 nt)} & \textbf{Mean Read Counts (\texttt{Input /\allowbreak{} Beads /\allowbreak{} HER2 /\allowbreak{} Myo /\allowbreak{} Thr})} & \textbf{R3 HER2 Raw $\log_2\text{FC}$} & \textbf{R3 Bead-Corrected $\log_2\text{FC}$} & \textbf{R3 HER2 $\text{FDR}$} & \textbf{Myoglobin \& $\alpha$-Thrombin Off-Target Enrichment} \\
\midrule
\endhead
\bottomrule
\endfoot
\textbf{\texttt{hybrid\_\allowbreak{}chimera\_\allowbreak{}008}} & \textbf{Round 3} (\texttt{8-\allowbreak{}bp stem + G4 chimera}) & \texttt{\seqsplit{GACGCCAATGGGTGGGGTTGTTTAGGGAGGGATTGGCGTC}} & \texttt{233 /\allowbreak{} 294 /\allowbreak{} 7,214 /\allowbreak{} 223 /\allowbreak{} 804} & $+4.95$ & \textbf{$+4.59$} & $0.031$ & No significant enrichment (Myo $+0.25$, Thr $+0.34$) \\
\addlinespace[4pt]
\arrayrulecolor{gray!35}\midrule[0.3pt]\arrayrulecolor{black}
\textbf{\texttt{constrained\_\allowbreak{}gquad\_\allowbreak{}005}} & \textbf{Round 3} (\texttt{8-\allowbreak{}bp stem + G4 chimera}) & \texttt{\seqsplit{CGCCACCATTAGGGTGGGAATAGGGAGGGAATTGGTGGCG}} & \texttt{242 /\allowbreak{} 177 /\allowbreak{} 4,773 /\allowbreak{} 207 /\allowbreak{} 157} & $+4.30$ & \textbf{$+4.47$} & $0.034$ & No significant enrichment (Myo $+0.43$, Thr $-0.18$) \\
\addlinespace[4pt]
\arrayrulecolor{gray!35}\midrule[0.3pt]\arrayrulecolor{black}
\textbf{\texttt{bef5bce6520a} ($\star$)} & \textbf{Round 2 $\rightarrow$ Round 3 carryover} (\texttt{6-\allowbreak{}bp stem + G4}) & \texttt{\seqsplit{AAGAGCGGGGTGGGTTTTCTTTGGGTTGGGCCGCTCAAAA}} & \texttt{258 /\allowbreak{} 504 /\allowbreak{} 8,657 /\allowbreak{} 285 /\allowbreak{} 311} & $+5.06$ & \textbf{$+4.28$} & $0.031$ & No significant enrichment (Myo $+0.18$, Thr $-0.69$) \\
\addlinespace[4pt]
\arrayrulecolor{gray!35}\midrule[0.3pt]\arrayrulecolor{black}
\textbf{\texttt{R1\_\allowbreak{}0130}} & \textbf{Round 1 $\rightarrow$ Round 3 carryover} (\texttt{G4-\allowbreak{}-\allowbreak{}hairpin hybrid}) & \texttt{\seqsplit{GGTGGTGGGTTGTGGGTGGTTGATTTTTCTAATGGTCATGAG}} & \texttt{343 /\allowbreak{} 314 /\allowbreak{} 4,747 /\allowbreak{} 208 /\allowbreak{} 560} & $+3.79$ & \textbf{$+3.74$} & $0.037$ & No significant enrichment (Myo $-0.21$, Thr $+0.69$) \\
\addlinespace[4pt]
\arrayrulecolor{gray!35}\midrule[0.3pt]\arrayrulecolor{black}
\textbf{\texttt{constrained\_\allowbreak{}gquad\_\allowbreak{}009}} & \textbf{Round 3} (\texttt{8-\allowbreak{}bp GC stem + G4 chimera}) & \texttt{\seqsplit{GCTCGCGCGGGTAAGGGTTTTTTGGGTAAGGGGCGCGAGC}} & \texttt{160 /\allowbreak{} 273 /\allowbreak{} 2,645 /\allowbreak{} 187 /\allowbreak{} 244} & $+4.04$ & \textbf{$+3.39$} & $0.034$ & No significant enrichment (Myo $+0.35$, Thr $-0.16$) \\
\addlinespace[4pt]
\arrayrulecolor{gray!35}\midrule[0.3pt]\arrayrulecolor{black}
\textbf{\texttt{constrained\_\allowbreak{}gquad\_\allowbreak{}002\_\allowbreak{}v3}} & \textbf{Round 3} (\texttt{8-\allowbreak{}bp stem + G4 chimera}) & \texttt{\seqsplit{GACGACGAGGGTAAGGGTTTTTTGGGTAAGGGTCGTCGTC}} & \texttt{219 /\allowbreak{} 196 /\allowbreak{} 2,075 /\allowbreak{} 183 /\allowbreak{} 186} & $+3.24$ & \textbf{$+3.22$} & $0.043$ & No significant enrichment (Myo $+0.27$, Thr $-0.07$) \\
\end{longtable}
\endgroup

\begin{landscape}
\begingroup
\scriptsize
\setlength{\tabcolsep}{2.2pt}
\renewcommand{\arraystretch}{1.15}
\vspace{8pt}\noindent{\small\bfseries Supplementary Table S3 | Authoritative summary of library cohorts, counter-screens, statistical hit criteria, motif ancestry, and statistically significant on-target hit rates across Rounds 1--3\par}\vspace{-4pt}
\begin{longtable}{@{}>{\raggedright\arraybackslash}p{0.065\linewidth}>{\raggedright\arraybackslash}p{0.065\linewidth}>{\raggedright\arraybackslash}p{0.080\linewidth}>{\raggedright\arraybackslash}p{0.060\linewidth}>{\raggedright\arraybackslash}p{0.115\linewidth}>{\raggedright\arraybackslash}p{0.090\linewidth}>{\raggedright\arraybackslash}p{0.070\linewidth}>{\raggedright\arraybackslash}p{0.070\linewidth}>{\raggedright\arraybackslash}p{0.080\linewidth}>{\raggedright\arraybackslash}p{0.080\linewidth}>{\raggedright\arraybackslash}p{0.080\linewidth}>{\raggedright\arraybackslash}p{0.060\linewidth}@{}}
\toprule
\textbf{Round} & \textbf{Target Protein} & \textbf{Counter-Targets \& Matrix Controls Screened} & \textbf{Full Sequenced Pool ($N_{\text{pool}}$)} & \textbf{Cohort Breakdown (\texttt{New Target Designs} / \texttt{Off-\allowbreak{}Target Sub-\allowbreak{}Pools} / \texttt{Variants \& Tiers} / \texttt{Controls \& Spikes})} & \textbf{Statistical Hit Inclusion Rule} & \textbf{Raw Enriched ($\log_2\text{FC} \ge 1.0$)} & \textbf{Volcano Pre-Hits ($\log_2\text{FC} \ge 1.0$, $\text{FDR} < 0.05$)} & \textbf{Statistically Significant On-Target Hits: New Target-Designed Candidates ($n/N, \%$)} & \textbf{On-Target Hits: Strictly \textit{De Novo} Subset ($n/N, \%$)} & \textbf{On-Target Hits: Full Sequenced Pool ($n/N, \%$)} & \textbf{Negative / Scrambled Controls ($n/N, \%$)} \\
\midrule
\endhead
\bottomrule
\endfoot
\textbf{Round 1} & \textbf{$\alpha$-Thrombin} & Myoglobin, insulin, bare streptavidin beads & 642 & \texttt{167} Thr new / \texttt{475} off-target (\texttt{196} Myo, \texttt{279} Ins) / \texttt{0} var / \texttt{0} ext ctrl & \texttt{DESeq2} $\log_2\text{FC} \ge 1.0$, $\text{FDR} < 0.05$ vs. input + beads & 1 / 642 (0.2\%) & 0 / 642 (0.0\%) & \textbf{0 / 167 (0.0\%)} & \textbf{0 / 57 (0.0\%)} \textit{(0 / 110 lit)} & \textbf{0 / 642 (0.0\%)} & 0 / 279 Ins decoy (0.0\%) \\
\addlinespace[4pt]
\arrayrulecolor{gray!35}\midrule[0.3pt]\arrayrulecolor{black}
\textbf{Round 1} & \textbf{Myoglobin} & $\alpha$-Thrombin, insulin, bare streptavidin beads & 642 & \texttt{196} Myo new / \texttt{446} off-target (\texttt{167} Thr, \texttt{279} Ins) / \texttt{0} var / \texttt{0} ext ctrl & \texttt{DESeq2} $\log_2\text{FC} \ge 1.0$, $\text{FDR} < 0.05$ vs. input + beads \textit{(37 at $\text{FDR} < 0.05$; 24 also $\ge 1.0\ \log_2\text{FC}$)} & 42 / 642 (6.5\%) & 37 / 642 (5.8\%) \textit{(24 at $\ge 1.0$)} & \textbf{20 / 196 (10.2\%)} \textit{(18 at $\ge 1.0$)} & \textbf{19 / 171 (11.1\%)} \textit{(1 / 25 lit; 34 / 499 [6.8\%] full pool)} & \textbf{37 / 642 (5.8\%)} \textit{(24 / 642 [3.7\%] at $\ge 1.0$)} & 0 / 53 random (0.0\%) \\
\addlinespace[4pt]
\arrayrulecolor{gray!35}\midrule[0.3pt]\arrayrulecolor{black}
\textbf{Round 1b}* & \textbf{Myoglobin} \textit{(Single-target expansion)} & Bare streptavidin beads & 1,341 & \texttt{1,288} Myo new / \texttt{0} off-target / \texttt{0} var / \texttt{53} scrambled neg & $\log_2\text{FC} \ge 1.0$, $\text{FDR} < 0.05$, OLS bead residual $> +1.00$ & 166 / 1,341 (12.4\%) & 148 / 1,341 (11.0\%) & \textbf{86 / 1,288 (6.7\%)} & \textbf{86 / 1,288 (6.7\%)} & \textbf{86 / 1,341 (6.4\%)} & 0 / 53 (0.0\%) \\
\addlinespace[4pt]
\arrayrulecolor{gray!35}\midrule[0.3pt]\arrayrulecolor{black}
\textbf{Round 2} & \textbf{$\alpha$-Thrombin} & Myoglobin, HER2, TNF-$\alpha$, bare streptavidin beads & 981 & \texttt{208} Thr new / \texttt{683} off-target (\texttt{243} Myo, \texttt{229} HER2, \texttt{211} TNF) / \texttt{0} var / \texttt{90} ctrls (\texttt{14} lit, \texttt{16} R1 spikes, \texttt{60} scrambles) & $\log_2\text{FC} \ge 1.0$, $\text{FDR} < 0.05$, OLS bead residual $> +1.00$, $x_i < 0.5$ \textit{(min hit res $+1.15$)} & 212 / 981 (21.6\%) \textit{(112 / 211 [53.1\%] vs. beads)} & 161 / 981 (16.4\%) \textit{(125 / 211 [59.2\%] Thr cohort)} & \textbf{54 / 208 (26.0\%)} \textit{(56 / 211 [26.5\%] w/ 2 lit ctrls)} & \textbf{10 / 76 (13.2\%)} \textit{(44 / 132 [33.3\%] lit)} & \textbf{73 / 981 (7.4\%)} \textit{(71 / 891 [8.0\%] non-ctrl)} & 0 / 60 (0.0\%) \\
\addlinespace[4pt]
\arrayrulecolor{gray!35}\midrule[0.3pt]\arrayrulecolor{black}
\textbf{Round 2} & \textbf{Myoglobin} & $\alpha$-Thrombin, HER2, TNF-$\alpha$, bare streptavidin beads & 977 & \texttt{244} Myo new / \texttt{643} off-target (\texttt{208} Thr, \texttt{224} HER2, \texttt{211} TNF) / \texttt{0} var / \texttt{90} ctrls (\texttt{14} lit, \texttt{16} R1 spikes, \texttt{60} scrambles) & $\log_2\text{FC} \ge 1.0$, $\text{FDR} < 0.05$, OLS bead residual $> +1.00$ & 177 / 977 (18.1\%) \textit{(12.4\% in R2 sub-analysis)} & 150 / 977 (15.4\%) \textit{(28 / 273 [10.3\%] Myo cohort)} & \textbf{9 / 244 (3.7\%)} \textit{(new Myo designs; 17 / 270 [6.3\%] incl. 8 / 16 R1 pos spikes; 17 / 183 [9.3\%] in $N=700$ benchmark)} & \textbf{9 / 238 (3.8\%)} \textit{(0 / 6 lit)} & \textbf{96 / 977 (9.8\%)} \textit{(88 / 977 [9.0\%] excl. 8 R1 pos spikes; 86 / 887 [9.7\%] non-ctrl)} & 1 / 60 (1.7\%) \textit{(0 / 53 primary random)} \\
\addlinespace[4pt]
\arrayrulecolor{gray!35}\midrule[0.3pt]\arrayrulecolor{black}
\textbf{Round 2} & \textbf{HER2 ECD} \textit{(Counter-screen)} & $\alpha$-Thrombin, myoglobin, TNF-$\alpha$, bare streptavidin beads & 968 & \texttt{229} HER2 new / \texttt{675} off-target / \texttt{0} var / \texttt{64} ctrls & $\log_2\text{FC} \ge 1.0$, $\text{FDR} < 0.05$, OLS bead residual $> +1.00$, non-binding on Myo \& Thr & 51 / 968 (5.3\%) & 46 / 968 (4.8\%) & \textbf{59 / 968 (6.1\%)} \textit{(all-pool bead-gated)} & \textbf{13 / 968 (1.3\%)} \textit{(strictly HER2-selective)} & \textbf{13 / 968 (1.3\%)} \textit{(strictly selective)} & 0 / 53 (0.0\%) \\
\addlinespace[4pt]
\arrayrulecolor{gray!35}\midrule[0.3pt]\arrayrulecolor{black}
\textbf{Round 3} & \textbf{$\alpha$-Thrombin} & Myoglobin, HER2, bare streptavidin beads & 1,441 \textit{(1,405 dedup)} & \texttt{311} Thr new / \texttt{749} off-target (\texttt{549} Myo, \texttt{200} HER2) / \texttt{148} var \& tiers (\texttt{88} mut, \texttt{60} tiers) / \texttt{197} ctrls (\texttt{84} pos spikes, \texttt{11} lit, \texttt{4} struct, \texttt{98} neg/rand) & $\log_2\text{FC} \ge 1.0$, $\text{FDR} < 0.05$, OLS bead residual $> +1.00$ (+ $y_{i,\text{thr}} - x_i > 1.0$ if bead pre-hit; \textit{min hit res $+1.94$}) & 394 / 1,441 (27.3\%) \textit{(251 / 311 [80.7\%] Thr new)} & 301 / 1,441 (20.9\%) \textit{(162 / 311 [52.1\%] Thr new)} & \textbf{131 / 311 (42.1\%)} & \textbf{106 / 286 (37.1\%)} \textit{(25 / 25 [100\%] lit)} & \textbf{162 / 1,405 (11.5\%)} \textit{(218 / 1,441 [15.1\%] sequenced)} & 0 / 20 scrambled/neg (0.0\%) \\
\addlinespace[4pt]
\arrayrulecolor{gray!35}\midrule[0.3pt]\arrayrulecolor{black}
\textbf{Round 3} & \textbf{Myoglobin} & $\alpha$-Thrombin, HER2, bare streptavidin beads & 1,441 \textit{(1,405 dedup)} & \texttt{549} Myo new / \texttt{511} off-target (\texttt{311} Thr, \texttt{200} HER2) / \texttt{148} var \& tiers (\texttt{88} mut, \texttt{60} tiers) / \texttt{197} ctrls (\texttt{84} pos spikes, \texttt{11} lit, \texttt{4} struct, \texttt{98} neg/rand) & $\log_2\text{FC} \ge 1.0$, $\text{FDR} < 0.05$, OLS bead residual $> +1.00$ & 369 / 1,441 (25.6\%) \textit{(164 / 549 [29.9\%] Myo new)} & 357 / 1,441 (24.8\%) \textit{(242 / 1,405 [17.2\%] dedup)} & \textbf{184 / 549 (33.5\%)} & \textbf{184 / 547 (33.6\%)} \textit{(0 / 2 [0.0\%] lit)} & \textbf{273 / 1,405 (19.4\%)} \textit{(292 / 1,441 [20.3\%] across replicates in Fig. 2D)} & 0 / 78 random/scrambled (0.0\%) \\
\addlinespace[4pt]
\arrayrulecolor{gray!35}\midrule[0.3pt]\arrayrulecolor{black}
\textbf{Round 3} & \textbf{HER2 ECD} & $\alpha$-Thrombin, myoglobin, bare streptavidin beads & 1,441 \textit{(1,405 dedup)} & \texttt{200} HER2 new / \texttt{860} off-target (\texttt{549} Myo, \texttt{311} Thr) / \texttt{148} var \& tiers / \texttt{197} ctrls & $\log_2\text{FC} \ge 1.0$, $\text{FDR} < 0.05$, OLS bead residual $> +1.00$, non-binding on Myo \& Thr & 396 / 1,441 (27.5\%) & 390 / 1,441 (27.1\%) & \textbf{8 / 200 (4.0\%)} \textit{(all-HER2 bead-gated)} & \textbf{8 / 200 (4.0\%)} & \textbf{71 / 1,405 (5.1\%)} \textit{(strictly HER2-selective; 117 / 1,405 [8.3\%] before cross-target filter)} & 0 / 24 (0.0\%) \\
\end{longtable}
\vspace{2pt}\noindent{\scriptsize \textit{*Footnote}: Summing the three focal closed-loop pools plotted in \textbf{Figures 2--3} (\texttt{Round 1} [$N = 642$], \texttt{Round 2} [$N = 981$], and \texttt{Round 3} [$N = 1,441$]) yields \textbf{3,064 sequenced library entries} (2,937 unique sequences after deduplicating carried-over controls). Including the supplementary single-target myoglobin expansion screen (\texttt{Round 1b}, $N = 1,341$ sequenced / $1,342$ ordered) brings the campaign total across all four synthesized pools to \textbf{4,405 sequenced constructs} (4,425 ordered constructs; \textbf{4,279 unique sequences} after cross-pool control deduplication, referred to in the text as $\sim 4,400$). "Literature Motif Descendants" (\texttt{lit}) denotes candidate sequences containing canonical $\alpha$-thrombin aptamer cores (\texttt{TBA} 15-nt \texttt{GGTTGGTGTGGTTGG} or \texttt{GGTTGGN\{2,4\}GGTTGG}, \texttt{HD22} core \texttt{\seqsplit{GGTAGGGCAGGTTGGGGT}}, \texttt{NU172} 25-nt \texttt{\seqsplit{GCCTAGGTTGGGTAGGGTGGTGGCG}} or 15-nt core \texttt{GGTTGGGTAGGGTGG}, or \texttt{RE31} core); "Strictly \textit{De Novo} Subset" excludes all such sequences.\par}

\endgroup
\end{landscape}

\begin{landscape}
\begingroup
\scriptsize
\setlength{\tabcolsep}{2.5pt}
\renewcommand{\arraystretch}{1.15}
\vspace{8pt}\noindent{\small\bfseries Supplementary Table S4 | Single-clone magnetic-bead fluorescence titration data and preliminary 1:1 Langmuir binding fits for full-length $\alpha$-thrombin candidates (\texttt{T01}--\texttt{T08})\par}\vspace{-4pt}
\begin{longtable}{@{}>{\raggedright\arraybackslash}p{0.070\linewidth}>{\raggedright\arraybackslash}p{0.075\linewidth}>{\raggedright\arraybackslash}p{0.120\linewidth}>{\raggedright\arraybackslash}p{0.130\linewidth}>{\raggedright\arraybackslash}p{0.055\linewidth}>{\raggedright\arraybackslash}p{0.055\linewidth}>{\raggedright\arraybackslash}p{0.090\linewidth}>{\raggedright\arraybackslash}p{0.070\linewidth}>{\raggedright\arraybackslash}p{0.080\linewidth}>{\raggedright\arraybackslash}p{0.080\linewidth}>{\raggedright\arraybackslash}p{0.095\linewidth}@{}}
\toprule
\textbf{Candidate ID} & \textbf{Synthesis Well \& Tube Ref (\texttt{Mfg \#}, Plate Row)} & \textbf{Architectural Family \& Motif Origin} & \textbf{Core Variable Sequence (\texttt{5' $\rightarrow$ 3'}, flanked by constant primers in full-length \texttt{75-\allowbreak{}-\allowbreak{}85 nt} construct)} & \textbf{Full Construct Len (nt)} & \textbf{R3 Pooled SELEX ($\log_2\text{FC}$)} & \textbf{Raw RFU (\texttt{0, 1, 10, 100, 1000 nM})} & \textbf{Specific Signal at $1\ \mu\text{M}$ ($\Delta\text{RFU}$, Fold / Blank)} & \textbf{Preliminary Apparent $K_D$, Fixed-Blank Fit ($R^2$)} & \textbf{Preliminary Apparent $K_D$, Free-Baseline Fit ($R^2$)} & \textbf{Binding Classification} \\
\midrule
\endhead
\bottomrule
\endfoot
\textbf{\texttt{T01} (\texttt{R3\_\allowbreak{}095\_\allowbreak{}var3})} & \texttt{\#579}, Row A (\texttt{Tube T07}) & \textbf{\texttt{NU172}-derived bimodal G4--duplex} (\texttt{C10A} bulge edit + \texttt{9-\allowbreak{}bp} terminal stem) & \texttt{\seqsplit{TAACCAGCGACTAGGTTGGGTAGGGTGGTGGCGCTGGTTA}} & 85 & $+16.27$ & \texttt{296, 478, 888, 1586, 2121} & $+1{,}825$ ($7.2\times$) & \textbf{$24.6 \pm 8.2\text{ nM}$} ($0.980$) & $32.0 \pm 13.8\text{ nM}$ ($0.986$) & \textbf{Strong nanomolar binder} \\
\addlinespace[4pt]
\arrayrulecolor{gray!35}\midrule[0.3pt]\arrayrulecolor{black}
\textbf{\texttt{T02}} & \texttt{\#325}, Row H (\texttt{Tube T02}) & \textbf{\texttt{TBA}-derived coaxial G4} (\texttt{15-\allowbreak{}nt} canonical \texttt{TBA} loop \texttt{GGTTGGTGTGGTTGG} + \texttt{11-\allowbreak{}bp} stem, $\Delta G = -12.0\text{ kcal/mol}$) & \texttt{\seqsplit{AACACGCCTATAGGTTGGTGTGGTTGGATAGGCGTGTT}} & 83 & $+11.88$ & \texttt{276, 278, 331, 702, 812} & $+536$ ($2.9\times$) & \textbf{$47.4 \pm 14.6\text{ nM}$} ($0.986$) & $43.4 \pm 16.4\text{ nM}$ ($0.989$) & \textbf{Strong nanomolar binder} \\
\addlinespace[4pt]
\arrayrulecolor{gray!35}\midrule[0.3pt]\arrayrulecolor{black}
\textbf{\texttt{T03}} & \texttt{\#342}, Row B (\texttt{Tube T01}) & \textbf{\texttt{TBA}-derived coaxial G4} (\texttt{15-\allowbreak{}nt} canonical \texttt{TBA} loop \texttt{GGTTGGTGTGGTTGG} + \texttt{8-\allowbreak{}bp} stem, $\Delta G = -8.2\text{ kcal/mol}$) & \texttt{\seqsplit{AGCTGACAGGTTGGTGTGGTTGGTGTCAGCT}} & 76 & $+11.89$ & \texttt{218, 306, 312, 337, 427} & $+209$ ($2.0\times$) & ---* & \textbf{$132.7\text{ nM}$} ($0.783$*) & \textbf{Moderate nanomolar binder} \\
\addlinespace[4pt]
\arrayrulecolor{gray!35}\midrule[0.3pt]\arrayrulecolor{black}
\textbf{\texttt{T04}} & \texttt{\#328}, Row F (\texttt{Tube T05}) & \textbf{\texttt{TBA}-derived loop-edited G4} (\texttt{11-\allowbreak{}bp} stem + central \texttt{TBA} loop \texttt{G8A} \texttt{TGT} $\rightarrow$ \texttt{TAT} specificity edit) & \texttt{\seqsplit{AAACACCGATAGGTTGGTATGGTTGGTATCGGTGTTT}} & 82 & $+9.81$ & \texttt{264, 283, 266, 335, 405} & $+141$ ($1.5\times$) & \textbf{$132.5 \pm 52.8\text{ nM}$} ($0.974$) & $146.0 \pm 77.6\text{ nM}$ ($0.975$) & \textbf{Moderate nanomolar binder} (reduced off-target enrichment in pooled screen) \\
\addlinespace[4pt]
\arrayrulecolor{gray!35}\midrule[0.3pt]\arrayrulecolor{black}
\textbf{\texttt{T05}} & \texttt{\#581}, Row C (\texttt{Tube T08}) & \textbf{Strictly \textit{de novo} hybrid G4--stem} (non-canonical G4 core + \texttt{10-\allowbreak{}bp} GC clamp) & \texttt{\seqsplit{CGGTGCTGCATGGGTGGGTAGGGTGGGACTTCCAGCACCG}} & 85 & $+3.00$ & \texttt{278, 269, 280, 262, 396} & $+118$ ($1.4\times$) & $> 1{,}000\text{ nM}$ & $> 1{,}000\text{ nM}$ & \textbf{Micromolar onset} (at $1\ \mu\text{M}$) \\
\addlinespace[4pt]
\arrayrulecolor{gray!35}\midrule[0.3pt]\arrayrulecolor{black}
\textbf{\texttt{T06}} & \texttt{\#912}, Row G (\texttt{Tube T06}) & \textbf{Strictly \textit{de novo} 3-tetrad G4} (3-plane G4 core + \texttt{6-\allowbreak{}bp} stem, $\Delta G = -6.9\text{ kcal/mol}$) & \texttt{\seqsplit{GCGCATGGGTTGGGATGGGTTGGGATGCGC}} & 75 & $+6.70$ & \texttt{291, 260, 268, 284, 349} & $+58$ ($1.2\times$) & $> 1{,}000\text{ nM}$ & $> 1{,}000\text{ nM}$ & \textbf{Trace micromolar onset} (at $1\ \mu\text{M}$) \\
\addlinespace[4pt]
\arrayrulecolor{gray!35}\midrule[0.3pt]\arrayrulecolor{black}
\textbf{\texttt{T07}} & \texttt{\#329}, Row D (\texttt{Tube T09}) & \textbf{Strictly \textit{de novo} 3-way junction} (non-G4 branched duplex) & \texttt{\seqsplit{CCGGCAGCTTGGTTGTGCGCCGGAAAAAATTCCGTGCCGG}} & 85 & $+2.41$ & \texttt{269, 261, 258, 277, 305} & $+36$ ($1.1\times$) & $> 1{,}000\text{ nM}$ (N.D.) & $> 1{,}000\text{ nM}$ (N.D.) & \textbf{Below $1\ \mu\text{M}$ detection ceiling} \\
\addlinespace[4pt]
\arrayrulecolor{gray!35}\midrule[0.3pt]\arrayrulecolor{black}
\textbf{\texttt{T08}} & \texttt{\#333}, Row E (\texttt{Tube T10}) & \textbf{Strictly \textit{de novo} C-rich hairpin} (non-G4 stem-loop) & \texttt{\seqsplit{CCTACCCTTCCTAACCCTAAGGTTCGGCACTGGCGA}} & 81 & $+3.29$ & \texttt{280, 247, 280, 276, 287} & $+7$ ($1.0\times$) & $> 1{,}000\text{ nM}$ (N.D.) & $> 1{,}000\text{ nM}$ (N.D.) & \textbf{Below $1\ \mu\text{M}$ detection ceiling} \\
\end{longtable}
\vspace{2pt}\noindent{\scriptsize \textit{Footnote on \texttt{T03} and $K_D$ fits: Candidates are indexed \texttt{T01}--\texttt{T08} in order of single-clone binding affinity, with synthesis manufacturing IDs (\texttt{Mfg \#}), 96-well plate rows (\texttt{A-\allowbreak{}-\allowbreak{}H}), and original synthesis tube labels (\texttt{Tube T01-\allowbreak{}-\allowbreak{}T10}; tubes \texttt{T03} and \texttt{T04} were not assayed on this plate) listed in Column 2 for cross-reference with raw plate-reader records. All fitted $K_D$ values should be considered preliminary estimates, as fluorescence signals did not reach a full saturation plateau ($B_{\max}$) within the $0\text{--}1{,}000\text{ nM}$ titration range. Anchoring $y_0 = 218\text{ RFU}$ distorts the low-concentration baseline for \texttt{T03} because its single $0\text{ nM}$ blank well (Row B) dipped to \texttt{218 RFU} (compared to the plate-wide blank mean of $271.5 \pm 23.9\text{ RFU}$ across rows A--H); the 3-parameter free-baseline hyperbolic fit ($\text{RFU}(C) = y_0 + \Delta B_{\max} \cdot C / (K_D + C)$) recovers a baseline of $y_0 = 271.7\text{ RFU}$ matching the plate blank and a preliminary apparent midpoint of $K_D = 132.7\text{ nM}$ ($\pm 227.5\text{ nM}$ standard error from 3-parameter covariance across 5 titration points; $R^2 = 0.783$) across the $0\text{--}1{,}000\text{ nM}$ range.}\par}

\endgroup
\end{landscape}

\end{document}